\documentclass[12pt]{article}

\usepackage{epsfig,latexsym,amsfonts,amsmath,amsthm,amssymb,amsbsy,multirow,slashed,wasysym,textcomp,dsfont,comment,mathtools,cancel,cite,diagbox,datetime,appendix,BOONDOX-calo}
\usepackage{tocloft}
\usepackage{bbold}
\usepackage{sidecap}
\usepackage{adjustbox}
\usepackage{pgfplots}
\usepackage{oplotsymbl}
\usepackage{tikzpagenodes}
\usepackage{adforn}
\usepackage{tikz}
\usetikzlibrary{backgrounds}
\usetikzlibrary{patterns}
\usetikzlibrary{arrows.meta}
\usetikzlibrary{shapes,decorations}
\tikzstyle{bag} = [align=center]
\usetikzlibrary{decorations.pathmorphing}
\usepackage[normalem]{ulem}
\usepackage[mathscr]{eucal}

\usepackage{hyperref}
\usepackage{subcaption}

 \newcommand{\badat}{\begin{alignedat}}
 \newcommand{\eadat}{\end{alignedat}}
 \newcommand\scalemath[2]{\scalebox{#1}{\mbox{\ensuremath{\displaystyle #2}}}}
 \def\be{\begin{equation}}
\def\ee{\end{equation}}

\def\p{\partial}

\usepackage{color}

\newcommand{\pink}[1]{\textcolor{\pink}{#1}}

\definecolor{dred}{rgb}{0.65,0.10,0.20} 
\definecolor{dblue}{rgb}{0.2,0.50,0.80}

\def\n{k}

\def\C{\mathcal{C}}

\def\J{\mathcal{J}}

\def\N{\mathcal{N}}
\def\O{\mathcal{O}}
\def\P{\mathcal{P}}

\def\S{\mathcal{S}}
\def\T{\mathcal{T}}

\def\Y{\mathcal{Y}}

\def\bh{{\bar h}}
\def\bz{{\bar z}}
\def\bw{{\bar w}}

\def\bk{{\bar k}}

\def\tA{\widetilde A}

\def\th{\widetilde h}

\def\th{\widetilde h}

\def\y{\mathcal{y}}

\def\zb{\bar{z}}
\def\wb{\bar{w}}

\def\bh{{\bar h}}
\def\bz{{\bar z}}
\def\bw{{\bar w}}

\def\Memo{{\rm M}}
\def\Gold{{\rm G}}
\def\CS{{\rm CS}}
\def\memo{{\rm m}}
\def\gold{{\rm g}}

\numberwithin{equation}{section} 
\pgfplotsset{compat=1.17} 

\begin{document}

 \begin{titlepage}
  \thispagestyle{empty}
  \begin{flushright}
  CPHT-RR037.052021
  \end{flushright}
  \bigskip
  \begin{center}

 \baselineskip=13pt {\LARGE \scshape{ Revisiting the Conformally Soft Sector\\[10pt] with Celestial Diamonds
}}

 \vskip1cm 

   \centerline{ 
   {Sabrina Pasterski}${}^\diamondsuit{}$,
   {Andrea Puhm}${}^\blacklozenge{}$,
   {and Emilio Trevisani}${}^\blacklozenge{}$
   }

\bigskip\bigskip
 
 \centerline{\em${}^\diamondsuit$ Princeton Center for Theoretical Science, Princeton, NJ 08544, USA}
 
\bigskip
 
\centerline{\em${}^\blacklozenge$   CPHT, CNRS, Ecole Polytechnique, IP Paris, F-91128 Palaiseau, France}

\smallskip

\bigskip\bigskip

\end{center}

\begin{abstract}
  \noindent

Celestial diamonds encode the structure of global conformal multiplets in 2D celestial CFT and offer a natural language for describing the conformally soft sector. The operators appearing at their left and right corners give rise to conformally soft factorization theorems, the bottom corners correspond to conserved charges, and the top corners to conformal dressings. 
We show that conformally soft charges can be expressed in terms of light ray integrals that select modes of the appropriate conformal weights.  They reside at the bottom corners of memory diamonds, and ascend to generalized currents. We then identify the top corners of the associated Goldstone diamonds with conformal Faddeev-Kulish dressings and compute the sub-leading conformally soft dressings in gauge theory and gravity which are important for finding nontrivial central extensions. Finally, we combine these ingredients to speculate on 2D effective descriptions for the conformally soft sector of celestial CFT.

\end{abstract}

\end{titlepage}

\setcounter{tocdepth}{2}
{\small \tableofcontents}

\section{Introduction}

The basic observable of quantum gravity in asymptotically flat spacetimes is the $S$-matrix. Celestial Holography purports a duality between gravitational scattering and a codimension~two conformal field theory (CFT) living on the celestial sphere, where the bulk Lorentz group induces global conformal transformations. When recast in a basis of boost eigenstates, scattering amplitudes transform as conformal correlators of primary operators in the dual celestial CFT. 

From the CFT perspective 2D is special: the global conformal group gets enhanced to local conformal symmetries. Remarkably, this infinite dimensional enhancement also appears in the 4D $S$-matrix. Even more surprisingly, the symmetry structure is much larger -- every soft factorization theorem gives a dual `current'. In particular, in addition to the enhancement of the 4D Lorentz group to local conformal transformations which are dual to Virasoro superrotations, translations get enhanced to BMS supertranslations. We thus expect the quantum gravity $S$-matrix to be highly constrained. Yet how all these constraints are organized in the 2D CFT dual is still an open question. Of course, the underlying motivation is that with a consistent 2D dual in hand, we hope to learn something interesting about quantum gravity in the bulk. A tipping point in the celestial holography paradigm will occur once an intrinsic description of celestial CFTs is achieved.

A first milestone is to find an intrinsic 2D CFT model of the spontaneous (asymptotic) symmetry breaking dynamics which is captured by the so-called {\it conformally soft} sector. In celestial CFTs the basic building blocks are conformal primary wavefunctions which have definite conformal dimension $\Delta$ and spin $J$ under the SL(2,$\mathbb{C}$) Lorentz group. They serve as the asymptotic states for celestial amplitudes and, for massless particles, can be written in terms of a Mellin transform in the energy of the usual plane wave scattering basis. These 4D wavefunctions define 2D primary operators in celestial CFT. While a $\delta$-function normalizable basis of conformal primary wavefunctions is obtained~\cite{Pasterski:2017ylz} for $\Delta \in 1+i\mathbb{R}$, the conformally soft limit involves an analytic continuation~\cite{Donnay:2018neh,Pasterski:2020pdk,Pasterski:2020pdk} to $\Delta \in \frac{1}{2}\mathbb{Z}$. 
Primary operators with conformally soft values of the dimension were shown to generate asymptotic symmetries in gauge theory and gravity. For example, spin-1 primary operators with $\Delta=1$ generate a U(1) Kac-Moody symmetry~\cite{Donnay:2018neh}, while BMS supertranslations and superrotations are generated by spin-2 operators picked out by taking $\Delta=1$ and 0, respectively~\cite{Donnay:2018neh,Pasterski:2020pdk}.

With these ingredients in hand, certain elements of an intrinsic 2D description have been explored. From the point of view of a hyperbolic foliation of Minkowski space~\cite{deBoer:2003vf}, these have a natural AdS$_3$/CFT$_2$ flavor~\cite{Cheung:2016iub}. The Kac-Moody symmetry associated to the leading soft theorem in gauge theory has been shown to arise from a Chern-Simons theory~\cite{Cheung:2016iub}. Meanwhile, the sub-leading soft graviton can be described by an Alekseev-Shatashvili action which governs the spontaneous breaking of Diff($S^2$) to Virasoro superrotations~\cite{Haehl:2019eae,Nguyen:2020hot,Nguyen:2021dpa}, giving new insight into the extensions of the BMS group~\cite{Campiglia:2014yka,Compere:2018ylh,Donnay:2020guq}.

This structure should generalize to encompass all conformally soft theorems (which occur for~$\lfloor s\rfloor +1$ distinct values of the conformal dimension whenever there is a massless particle with bulk spin~$s$). In~\cite{Pasterski:2021fjn}, we showed that celestial CFT primaries and their descendants organize into `celestial diamonds' which capture the conformally soft physics. The wavefunction based approach employed there exploited the power of the embedding space formalism applied to 2D celestial CFT. Here, we push these results forward from statements about wavefunctions to statements about 2D (and 4D) operators. This lets us appreciate the extrapolate dictionary in asymptotically flat spacetimes, understand that soft charges are celestial primary descendants,\footnote{This part is very much inspired by the interesting recent work of Banerjee et al.~\cite{Banerjee:2018gce, Banerjee:2018fgd,Banerjee:2019aoy,Banerjee:2019tam}.} identify conformal dressings for celestial amplitudes, and speculate about intrinsic 2D descriptions of celestial CFT. This approach merges and extends the soft charge analysis of~\cite{Banerjee:2019aoy,Banerjee:2019tam} and the leading conformally soft Faddeev-Kulish dressings of~\cite{Arkani-Hamed:2020gyp}. For concreteness we focus on the leading and sub-leading celestial diamonds in gauge theory and gravity.
\pagebreak 

Our main results are as follows. Formulating an extrapolate-style dictionary for celestial CFTs, we find that states of definite $(\Delta,J)$ are prepared with bulk operators integrated along light rays and pushed to the conformal boundary. For conformally soft values of the dimension, these light ray integrals pick out soft charges for spontaneously broken asymptotic symmetries. Recasting them as celestial sphere integrals of conformally soft operators, we see that the latter correspond to primary descendants residing at the bottom corners of celestial diamonds associated to electromagnetic and gravitational memory. Generalized celestial currents living on the edge of celestial memory diamonds can be understood as `ascendants' of these conformally soft charge operators. This includes the BMS supertranslation current~\cite{Strominger:2013jfa} whose operator product involves a shift $\Delta \to \Delta+1$ that presents one of the more exotic features of celestial CFT~\cite{Donnay:2018neh}.

Celestial diamonds for Goldstone operators provide conformal dressings for celestial amplitudes. The leading Faddeev-Kulish dressings in gauge theory and gravity rendering scattering amplitudes infrared finite were shown in~\cite{Arkani-Hamed:2020gyp} to be given by Goldstone boson insertions in the conformal basis. In the language of celestial diamonds these correspond to the operators at the top corners which are associated to generalized ($s\leq |J|$) rather than radiative ($s=\pm J$) conformal primary wavefunctions. Here we extend this result to sub-leading conformally soft dressings building on the results of~\cite{Choi:2019rlz}. In particular, we identify the sub-leading conformally soft Faddeev-Kulish dressing in gravity as arising from the operator at the top corner of the sub-leading graviton Goldstone diamond that descends to the dual stress tensor of~\cite{Ball:2019atb}.  This operator, as well as the celestial stress tensor, give rise to two point functions which can encode non-trivial central charges.

The operators at the top corners of celestial Goldstone and memory diamonds govern the spontaneous symmetry breaking dynamics. This should be described by an intrinsically 2D effective theory. We propose a higher derivative Gaussian model that captures the free limit of the aforementioned examples proposed in the literature. Our toy model reproduces several features of celestial diamonds, such as the shortening conditions of conformal multiplets and the appearance of generalized celestial currents living at their edges.

This paper is organized as follows.  We review the 4D bulk wavefunctions relevant for 2D celestial CFT operators in section~\ref{sec:BulktoBdyOps}. In section~\ref{sec:softcharges} we show that conformally soft charges at leading and sub-leading order in gauge theory and gravity correspond to conformal primary descendants at the bottom corners of the corresponding celestial memory diamonds. We discuss how they are selected by light ray integrals of general bulk fields pushed to the boundary. We then turn to the Goldstone diamonds in section~\ref{sec:dressing}, where we identify the top corners with conformal Faddeev-Kulish dressings. In section~\ref{sec:conf_soft_sec}, we combine these ingredients to model the 2D effective theory.

\section{From Wavefunctions to Operators}
\label{sec:BulktoBdyOps}
In section~\ref{sec:CPWs} we review the construction of conformal primary wavefunctions on $\mathbb{R}^{1,3}$ and then use them to create local operators on the celestial sphere in section~\ref{sec:BndyOp}.

\subsection{Bulk Wavefunctions}
\label{sec:CPWs}

Let us start with the definition of conformal primary wavefunctions. These are functions of a bulk point $X\in\mathbb{R}^{1,3}$ and a boundary point $(w,\bw)\in \mathbb{C}$ on the celestial sphere.  We will focus on massless fields here. 
\vspace{1em}

\noindent{\bf Definition:} A {\it conformal primary wavefunction} is a function on $\mathbb{R}^{1,3}$ which transforms under SL(2,$\mathbb{C}$) as a 2D conformal primary of conformal dimension $\Delta$ and spin $J$, and a 4D (spinor-) tensor field of spin-$s$, namely:
\begin{equation}\label{Defgenprim}
    \badat{2}
\Phi^{s}_{\Delta,J}\Big(\Lambda^{\mu}_{~\nu} X^\nu;\frac{a w+b}{cw+d},\frac{{\bar a} \bw+{\bar b}}{{\bar c}\bw+{\bar d}}\Big)=(cw+d)^{\Delta+J}({\bar c}\bw+{\bar d})^{\Delta-J}D_s(\Lambda)\Phi^{gen,s}_{\Delta,J}(X^\mu;w,\bw)\,,
\eadat
\end{equation}
where $D_s(\Lambda)$ is the 3+1D spin-$s$ representation of the Lorentz algebra. \vspace{1em}

\noindent We call such a wavefunction a {\it radiative} conformal primary if $s=|J|$ and it satisfies the linearized equations of motion. Meanwhile off-shell wavefunctions and those with $|J|<s$ go under the umbrella of {\it generalized} conformal primaries. 

\subsubsection*{From Planes Waves to Conformal Primary Wavefunctions}

A point $(w,\bw)$ on the celestial sphere can be embedded into the null cone of $\mathbb{R}^{1,3}$ as follows
\be\label{qmu}
q^\mu=(1+w\bw,w+\bw,i(\bw-w),1-w\bw)\,.
\ee
From this we obtain two natural polarization vectors $\sqrt{2}\epsilon_w^\mu=\p_w q^\mu$ and $\sqrt{2}\epsilon_\bw^\mu=\p_\bw q^\mu$.
Massless spin-$s$ particles are labeled by their null momentum four vectors $k^\mu=\omega q^\mu(w,\bw)$ and polarization. 
We can map these momentum eigenstates to boost eigenstates via a Mellin transform. For a function $f(\omega)$ the Mellin transform is defined by
\begin{equation}\label{Mellin}
\mathcal{M}[f](\Delta)=\int_0^\infty d\omega \omega^{\Delta-1}f(\omega)\equiv \phi(\Delta)\,,
\end{equation}
with the inverse transform given by
\begin{equation}\label{InverseMellin}
\mathcal{M}^{-1}[\phi](\omega)=\frac{1}{2\pi i}\int_{c-i\infty}^{c+i\infty} d\Delta \, \omega^{-\Delta}\phi(\Delta)=f(\omega)\,,
\end{equation}
where, for our purposes, we can take the contour $c=1$~\cite{Donnay:2020guq}. 
For a scalar plane wave, a Mellin transform in the energy $\omega$ takes us to
\be\label{scalar1}
\int_0^\infty d\omega \omega^{\Delta-1}e^{\pm i\omega q\cdot X_\pm}=\frac{\Gamma(\Delta)}{(\pm i)^\Delta}\frac{1}{(-q\cdot X_\pm)^\Delta}\, , 
\ee
which has conformal dimension $\Delta$ and spin $J=0$. 
Here $X^\mu_\pm=X^\mu \mp i\varepsilon(1,0,0,0)$ is a regulator which makes this integral converge. We will suppress the $\pm$ superscripts unless necessary. 
Mellin transformed plane waves with $s>0$ are obtained by multiplying~\eqref{scalar1} with $\epsilon_w$ and $\epsilon_\bw$.  However, these are only gauge equivalent to conformal primaries~\eqref{Defgenprim}. To obtain wavepackets for \mbox{spin-$s$} particles that transform with definite $(\Delta,J)$ under an SL(2,$\mathbb{C}$) transformation we use the spacetime dependent polarization vectors
\be\label{tetrad}
m^\mu=\epsilon^\mu_w-\frac{\epsilon_w\cdot X}{q\cdot X}q^\mu\,, ~~~\bar{m}^\mu=\epsilon^\mu_\bw -\frac{\epsilon_\bw\cdot X}{q\cdot X}q^\mu\,,
\ee
which transform with conformal dimension $\Delta=0$ and spin $J=\pm1$.

The simplest $J=+s$ bosonic radiative conformal primary wavefunctions take the form
\be\label{PhisD}
\Phi^{s}_{\Delta,+s}=m_{\mu_1}...m_{\mu_s}\varphi^\Delta\,,
\ee
and similarly for $J=-s$ with $m\mapsto \bar{m}$, where we have defined the scalar conformal primary wavefunction without the normalization factor in~\eqref{scalar1} as
\be\label{scalar2}
\varphi^{\Delta,\pm}=\frac{1}{(-q\cdot X_\pm)^\Delta}\,.
\ee
Up to an overall normalization, these are gauge equivalent to Mellin representatives of the corresponding plane wave solutions, and we note in passing that the spin-0,~1 and 2 cases obey a classical Kerr-Schild double copy with $m^\mu$ and $\bar m^\mu$ corresponding to the Kerr-Schild vectors~\cite{Pasterski:2020pdk}.
Under a shadow transform the wavefunctions in~\eqref{PhisD} get mapped to
\begin{equation}\label{ShadowPhi}
\widetilde{\Phi}^s_{\Delta,J}=\widetilde{{\Phi}^s_{2-\Delta,-J}}\,,
\end{equation}
where \be\label{PhisDsh}
\tilde{\Phi}^{s}_{\Delta,+s}=(-1)^s(-X^2)^{\Delta-1}{ m}_{\mu_1}...{ m}_{\mu_s}\varphi^\Delta\,.
\ee
As shown in~\cite{Pasterski:2017kqt}, we can capture finite energy scattering states using~\eqref{PhisD} with conformal dimensions on the principal series 
\be
\Delta=1+i\lambda\,,~~~\lambda\in \mathbb{R}\,.
\ee
 An equivalent basis uses the same spectrum but takes a shadow transform of the reference direction.

\subsubsection*{Goldstone and Memory Wavefunctions}

Celestial amplitudes are obtained from ordinary amplitudes by Mellin transforming the energy of each external particle. They obey universal factorization theorems, called {\it conformally soft} theorems, when one of the conformal dimensions is taken to special values of $\Delta\in \mathbb{Z}$ (for integer bulk spin).  This requires us to analytically continue our radiative wavefunctions to  $\Delta\in \mathbb{C}$. We will now catalog the conformally soft modes used in this paper. 

For $J=+s=+1$ the modes which pick out the leading and sub-leading conformally soft theorems are given by
\begin{equation}\label{Aconfsoft}
    A^\Gold_{1,+1;\mu}=m_\mu \varphi^1\,, \quad A^\gold_{0,+1;\mu}=m_\mu\,, \quad \tA^\gold_{2,+1;\mu}=X^2 m_\mu \varphi^2\,,
\end{equation}
while for $J=+s=+2$ they are
\begin{equation}\label{hconfsoft}
    h^\Gold_{1,+2;\mu\nu}=m_\mu m_\nu \varphi^1\,, \quad h^\Gold_{0,+2;\mu\nu}=m_\mu m_\nu\,, \quad \th^\Gold_{2,+2;\mu\nu}=-X^2 m_\mu m_\nu\varphi^2\,,
\end{equation}
with similar expressions for $J=-s$ with $m \mapsto \bar m$. Here we have adopted the notation of~\cite{Pasterski:2021fjn}: the label $\Gold$ denotes Goldstone modes of spontaneously broken asymptotic symmetries associated to the leading soft photon theorem and the leading and sub-leading soft graviton theorem.  Meanwhile, the sub-leading soft photon theorem arises from conformal primaries which are not pure gauge. We denote these with the label $\gold$. 
\begin{figure}[hb!]
\centering
\begin{subfigure}{0.45\linewidth}
\centering
\begin{tikzpicture}[scale=.9]
\draw[thick,->,dred] (-1+.05,1-.05)node[left]{
} --node[below left]{
} (-.05,.05) ;
\draw[thick,->,dred] (1,1)-- (.05,.05);
\filldraw[dred] (0,0) circle (2pt);
\filldraw[dred] (-1,1) circle (2pt) ;
\draw[thick,dblue!50!white] (0+.1414/2,1+.1414/2) arc (45:-135:.1);
\filldraw[dred] (1,1) circle (2pt) ;
\filldraw[dred] (0,2) circle (2pt) ;
\draw[thick,->,dred] (0,2)-- (1-.05,1+.05);
\node at (-2,0) {$2$};
\node at (-2,1) {$1$};
\node at (-2,2) {$0$};
\node at (-1.6,3) {$J$};
\node at (-2,2.87) {$\tiny{\Delta}$};
\draw[black] (-2+.01,3+.2) -- (-2+.28,3-.3);
\node at (-1,3) {$-1$};
\node at (0,3) {$0$};
\node at (1,3) {$1$};
\filldraw[black,dblue] (1,0) circle (2pt) ;
\filldraw[black,dblue!50!white] (-1,0) circle (2pt) ;
\filldraw[black,dblue!50!white] (1,2) circle (2pt) ;
\filldraw[black,dblue] (-1,2) circle (2pt) ;
\draw[->,thick,dblue!50!white] (1,2) --  (0+.07,1+.07);
\draw[->,thick,dblue!50!white]  (0-.07,1-.07) -- (-1+.05,.05);
\draw[thick,dblue] (0-.1414,1+.1414) arc (135:315:.2);
\draw[->,thick,dblue] (-1,2) --  (0-.1414,1+.1414);
\draw[->,thick,dblue] (0+.1414,1-.1414) -- (1-.03,0.03);
\draw[thick,->,dred] (0,2)-- (-1+.05,1.05);
\filldraw[white] (0,0) circle (1pt) ;
\node[fill=dred,regular polygon, regular polygon sides=4,inner sep=1.6pt] at (0,2) {};
\node[fill=white,regular polygon, regular polygon sides=4,inner sep=.8pt] at (0,2) {};
\node at (-3,-1) {};
\node at (3,0) {};
\node at (2,4) {};
\end{tikzpicture}
\caption{}
\end{subfigure}
\begin{subfigure}{0.45\linewidth}
\centering
\begin{tikzpicture}[scale=.9]
\definecolor{mygray}{rgb}{.7, 0.7, .7};
\definecolor{mygray2}{rgb}{.4, 0.4, .4};
\filldraw[dred] (-2,1) circle (2pt) ;
\filldraw[dred] (2,1) circle (2pt) ;
\filldraw[dblue] (-2,2) circle (2pt) ;
\filldraw[dblue] (2,0) circle (2pt) ;
\filldraw[dblue] (1,-1) circle (2pt) ;
\filldraw[dblue!50!white] (2,2) circle (2pt) ;
\filldraw[dblue!50!white] (-2,0) circle (2pt) ;
\filldraw[dblue!50!white] (-1,-1) circle (2pt) ;
\draw[thick,->,dblue] (2,0)-- (1+.05,-1+.05);
\draw[thick,->,dblue] (-1,3)-- (-2+.05,2+.05);
\draw[thick,mygray] (0+.1414/2,1+.1414/2) arc (45:-135:.1);
\draw[thick,mygray] (1+.1414/2,2+.1414/2) arc (45:-135:.1);
\draw[thick,mygray] (-1+.1414/2,0+.1414/2) arc (45:-135:.1);
\draw[thick,dred] (1+.1414/2,0+.1414/2) arc (45:-135:.1);
\draw[thick,dred] (-1+.1414/2,2+.1414/2) arc (45:-135:.1);
\draw[thick,dblue!50!white] (0+.1414/2,0+.1414/2) arc (45:-135:.1);
\draw[thick,dblue!50!white] (1+.1414/2,1+.1414/2) arc (45:-135:.1);
\draw[thick,dblue!50!white] (-1+.1414/2,1+.1414/2) arc (45:-135:.1);
\draw[thick,dblue!50!white] (0+.1414/2,2+.1414/2) arc (45:-135:.1);
\draw[thick,dblue] (0-.1414,0+.1414) arc (135:315:.2);
\draw[thick,dblue] (1-.1414,1+.1414) arc (135:315:.2);
\draw[thick,dblue] (0-.1414,2+.1414) arc (135:315:.2);
\draw[thick,dblue] (-1-.1414,1+.1414) arc (135:315:.2);
\draw[->,thick,dblue] (-2,2) --  (-1-.1414,1+.1414);
\draw[->,thick,dblue] (-1,3) --  (0-.1414,2+.1414);
\draw[->,thick,dblue] (-1+.1414,1-.1414) --  (0-.1414,0+.1414);
\draw[->,thick,dblue] (0+.1414,0-.1414) -- (1-.05,-1+.05);
\draw[->,thick,dblue] (0+.1414,2-.1414) --  (1-.1414,1+.1414);
\draw[->,thick,dblue] (1+.1414,1-.1414) -- (2-.05,0+.05);
\draw[->,thick,dblue!50!white] (1,3) -- (2-.05,2+.05);
\draw[->,thick,dblue!50!white] (1,3) -- (0+.07,2+.07);
\draw[->,thick,dblue!50!white] (-2,0) -- (-1-.05,-1+.05);
\draw[->,thick,dblue!50!white] (2,2) -- (1+.07,1+.07);
\draw[->,thick,dblue!50!white] (1-.07,1-.07) -- (0+.07,0+.07);
\draw[->,thick,dblue!50!white] (0-.07,2-.07) -- (-1+.07,1+.07);
\draw[->,thick,dblue!50!white] (-1-.07,1-.07) -- (-2+.05,0+.05);
\draw[->,thick,dblue!50!white] (0-.07,0-.07) -- (-1+.05,-1+.05);
\draw[->,thick,dred] (0,3) --  (1-.1414,2+.1414);
\draw[->,thick,dred] (0,3) --  (-1+.07,2+.07);
\draw[->,thick,dred] (-1-.07,2-.07)-- (-2+.05,1+.05);
\draw[->,thick,dred] (1-.07,0-.07)-- (0+.05,-1+.05);
\filldraw[mygray2] (-2,3) circle (2pt) ;
\filldraw[mygray] (2,3) circle (2pt) ;
\filldraw[black,dred] (0,3) circle (2pt) ;
\filldraw[white] (-1,-1) circle (1pt) ;
\filldraw[dblue!50!white] (1,3) circle (2pt) ;
\filldraw[white] (1,-1) circle (1pt) ;
\filldraw[mygray2] (2,-1) circle (2pt) ;
\filldraw[dred] (0,-1) circle (2pt) ;
\filldraw[mygray] (-2,-1) circle (2pt) ;
\filldraw[white] (0,-1) circle (1pt) ;
\node at (-3,-1) {$3$};
\node at (-3,0) {$2$};
\node at (-3,1) {$1$};
\node at (-3,2) {$0$};
\node at (-3,3) {$-1~~$};
\node at (-2.6,4) {$J$};
\node at (-3,3.87) {$\tiny{\Delta}$};
\node at (-2,4) {$-2$};
\draw[thick] (-3+.01,4+.2) -- (-3+.28,4-.3);
\node at (-1,4) {$-1$};
\node at (0,4) {$0$};
\node at (1,4) {$1$};
\node at (2,4) {$2$};
\draw[->,thick,dred] (2,1) --  (1+.07,0+.07);
\draw[->,thick,mygray] (1-.07,2-.07) --  (0+.07,1+.07);
\draw[->,thick,mygray]  (0-.07,1-.07) -- (-1+.07,.07);
\draw[->,thick,mygray]  (2,3) -- (1+.07,2.07);
\draw[->,thick,mygray]  (-1-.07,0-.07) -- (-2+.05,-1+.05);
\draw[thick,mygray2] (0-.1414,1+.1414) arc (135:315:.2);
\draw[thick,mygray2] (-1-.1414,2+.1414) arc (135:315:.2);
\draw[thick,mygray2] (1-.1414,0+.1414) arc (135:315:.2);
\draw[->,thick,mygray2] (-1+.1414,2-.1414) --  (0-.1414,1+.1414);
\draw[->,thick,mygray2] (0+.1414,1-.1414) -- (1-.1414,.1414);
\draw[->,thick,mygray2] (1+.1414,0-.1414) -- (2-.05,-1+.05);
\draw[->,thick,mygray2] (-2,3) --  (-1-.1414,2+.1414);
\draw[->,thick,dred] (-2,1) --  (-1-.1414,0+.1414);
\draw[thick,dred] (-1-.1414,0+.1414) arc (135:315:.2);
\draw[->,thick,dred] (-1+.1414,0-.1414) -- (0-.05,-1+.05);
\draw[->,thick,dred] (1+.1414,2-.1414) -- (2-.05,1+.05);
\draw[thick,dred] (1-.1414,2+.1414) arc (135:315:.2);
\node[fill=dblue,regular polygon, regular polygon sides=4,inner sep=1.6pt] at (-1,3) {};
\node[fill=white,regular polygon, regular polygon sides=4,inner sep=.8pt] at (-1,3) {};
\node[fill=dred,regular polygon, regular polygon sides=4,inner sep=1.6pt] at (0,3) {};
\node[fill=white,regular polygon, regular polygon sides=4,inner sep=.8pt] at (0,3) {};
\node[fill=dblue!50!white,regular polygon, regular polygon sides=4,inner sep=1.6pt] at (1,3) {};
\node[fill=white,regular polygon, regular polygon sides=4,inner sep=.8pt] at (1,3) {};
\node at (3.5,0) {};
\end{tikzpicture}
\caption{}
\end{subfigure}
\caption{Celestial diamonds for the conformally soft (a) photons and (b) gravitons.
\label{celestialdiamonds}}
\end{figure}
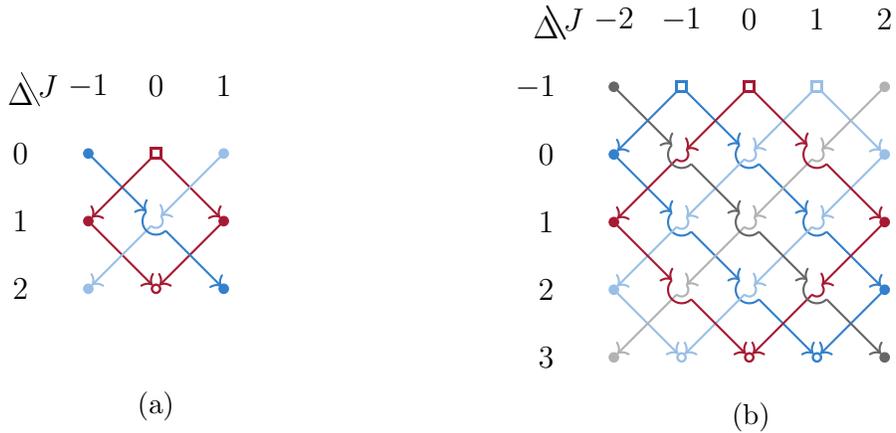

In~\cite{Pasterski:2021fjn} we showed that the radiative conformal primaries associated to non-trivial soft theorems are connected by descendancy relations to non-radiative generalized conformal primaries at the top and bottom corners. These are summarized for the spin-1 and spin-2 cases in figure~\ref{celestialdiamonds}. 
The leading soft theorems are shown in red, the sub-leading ones in blue and the subsub-leading ones in grey.
For each diamond drawn in figure~\ref{celestialdiamonds} there are actually two copies corresponding to symplectically paired modes: one associated to conformal soft theorems, the other to non-trivial memory effects.   In particular, we have for $J=+s=+1$
\begin{equation}\label{Aconfmem}
     A^\memo_{2,+1;\mu}=m_\mu \varphi^2\,, \quad \tA^\memo_{0,+1;\mu}=X^{-2} m_\mu \,,
\end{equation}
while for $J=+s=+2$ they are
\begin{equation}\label{hconfmem}
     h^\Memo_{2,+2;\mu\nu}=m_\mu m_\nu\varphi^2\,, \quad \th^\Memo_{0,+2;\mu\nu}=-X^{-2} m_\mu m_\nu\,.
\end{equation}
The case $\Delta=1$ is more subtle as the radiative wavefunction and its shadow degenerate~\cite{Donnay:2018neh}. The symplectic partners of the the $\Delta=1$ Goldstone modes in~\eqref{Aconfsoft} and~\eqref{hconfsoft} were constructed in~\cite{Donnay:2018neh} via a special limiting procedure to produce non-trivial memory effects. They are given by the $\Delta=1$ conformally soft photon and graviton wavefunctions
\begin{equation}\label{AhCS}
    A^{\CS}_{1,+1;\mu}=m_\mu \varphi^{\CS}\,,\quad  h^{\CS}_{1,+2;\mu\nu}=m_\mu m_\nu \varphi^{\CS}\,,
\end{equation}
and similarly for $J\mapsto -J$ and $m\mapsto \bar{m}$ where, as  in \cite{Pasterski:2020pdk}, we have defined the $\Delta=1$ conformally soft scalar wavefunction 
\begin{equation}\label{eq:phiCS} 
    \varphi^{\CS}=\left[\Theta(X^2)+{\rm log}(X^2)(q\cdot X)\delta(q\cdot X)\right]\varphi^1\,\equiv\varphi^{\CS'}+\varphi^{\CS''} .
\end{equation}
The $\CS'$ wavefunctions with $\varphi^{\CS'}\equiv \Theta(X^2)\varphi^1$ describe solutions that are glued across the lightcone (and will reappear in section~\ref{sec:dressing}) while the $\CS''$ wavefunctions with $\varphi^{\CS''}\equiv\log(X^2)\delta(q\cdot X)$ correspond to shockwave solutions~\cite{Pasterski:2020pdk}.

\subsection{Boundary Operators}
\label{sec:BndyOp}

Given a 4D bulk operator $O^{s}(X^\mu)$ of spin-$s$ in the Heisenberg picture, we can define a 2D operator in the celestial CFT by~\cite{Donnay:2020guq}
 \be\label{eq:2Dop}
\O^{s,\pm}_{\Delta,J}(w,\bw)\equiv i(O^{s}(X^\mu),\Phi^s_{\Delta^*,-J}(X_\mp^\mu;w,\bw))\,,
\ee
where the $\pm$ on the operator indicates whether it corresponds to an {\it in} or an {\it out} state and this formula uses standard inner products $(.\,,.)$ computed on a Cauchy slice in the bulk. 

\subsubsection*{Operators in the Bulk}

A natural inner product between complex spin-1 wavefunctions is given by
 \begin{equation}\label{eq:IPspin1}
 (A, A')_\Sigma =- i \int d\Sigma^\rho \,\left[ A^{\nu} {F'}_{\rho\nu}^{*} 
 -{A'}^{* \nu} F_{\rho\nu} 
 \right] \,,
 \end{equation}
and between complex spin-2 wavefunctions by (see e.g.~\cite{Ashtekar:1987tt,Crnkovic:1986ex,Lee:1990nz,Wald:1999wa}) \begin{equation}\label{eq:IPspin2}
( h,h')_\Sigma=-i\int d\Sigma^\rho \Big[ h^{\mu \nu} \nabla_\rho h'^{*}_{\,\,\mu \nu}-2h^{\mu \nu} \nabla_\mu h'^{*}_{\,\,\rho\nu}+h\nabla^\mu h'^{*}_{\,\,\rho\mu}-h \nabla_\rho h'^{*}+h_{\rho\mu} \nabla^\mu h'^{*} - (h \leftrightarrow h'^{*})\Big]\,,
\end{equation}
where $\Sigma$ is a Cauchy surface in the bulk, $h=h^\sigma_{\,\, \sigma}$ vanishes for radiative wavefunctions due to the tracelessness condition, and we make note of the complex conjugation in the definition.   These will be used within~\eqref{eq:2Dop}. If we want to express the operators~\eqref{eq:2Dop} in terms of the standard creation and annihilation operators, we will also need the mode expansion of the bulk field operators. In the momentum basis, we have
 \begin{equation}\label{eq:Aexp}
    \hat{A}_\mu(X)=e\int \frac{d^3k}{(2\pi)^3}\frac{1}{2k^0}\left[a_\mu e^{ik\cdot X}+a_\mu^\dagger e^{-ik\cdot X}\right]\,,
\end{equation}
for spin-1  and 
\begin{equation}\label{eq:hexp}
    \hat{h}_{\mu\nu}(X)=\kappa\int \frac{d^3k}{(2\pi)^3}\frac{1}{2k^0}\left[a_{\mu\nu} e^{i k\cdot X}+a_{\mu\nu}^\dagger e^{-ik\cdot X}\right]\,,
\end{equation}
for spin-2, where $\kappa=\sqrt{32\pi G}$.   Here we have defined
\be\label{eq:apolspin12}
a_\mu=\sum_{\alpha\in\pm} \epsilon_\mu^{\alpha *}a_\alpha,~~~a_{\mu\nu}=\sum_{\alpha\in\pm} \epsilon_{\mu\nu}^{\alpha *}a_\alpha\,,
\ee
where $\epsilon_{\mu\nu}^{\alpha*}=\epsilon_\mu^{\alpha*} \epsilon_\nu^{\alpha*}$, with $\alpha=w(\bw)$ corresponding to helicity $\alpha=+(-)$. We have left the spin~$s$ of the mode operators implicit since no confusion should arise. We can go from energies to conformal dimensions with a Mellin transform\footnote{Recall that $k^\mu=\omega q^\mu$. In order to go from round Bondi coordinates to a flat celestial sphere the parameter we Mellin transform is not the energy but rather $\omega=k^0/(1+w\bw)$~\cite{Donnay:2018neh}.
To compare to standard expressions~\cite{Weinberg:1995mt} one should keep in mind that we mean $a_\pm(E=\omega(1+w\bw);w,\bw)$.}
\be\label{mellinmode}
a_{\Delta,\pm s}(w,\bw)=\int_0^\infty d\omega \omega^{\Delta-1}a_\pm(\omega,w,\bw)\,,\quad a^\dagger_{\Delta,\mp s}(w,\bw)=\int_0^\infty d\omega \omega^{\Delta-1}a^\dagger_\pm(\omega,w,\bw)\,,
\ee
and back again to the momentum basis with an inverse Mellin transform
\begin{equation}\label{eq:inversemellinmode}
    a_\pm(\omega)=\frac{1}{2\pi} \int_{1-i\infty}^{1+i\infty} (-id\Delta) \omega^{-\Delta}a_{\Delta,\pm s}\,,\quad    a^\dagger_\mp(\omega)=\frac{1}{2\pi} \int_{1-i\infty}^{1+i\infty} (-id\Delta) \omega^{-\Delta}a^\dagger_{\Delta,\pm s}\,.
\end{equation}
Here and in what follows we label both the creation and annihilation Mellin operators with the $\Delta$ and $J$ of the corresponding state on the celestial sphere, i.e. by $a^\dagger_{\Delta,J}$ we will mean $(a^\dagger)_{\Delta,J}$ versus $(a_{\Delta,J})^\dagger$.  This matches the helicity after crossing to an all-{\it out} configuration. 

\subsubsection*{Operators at the Boundary}

The 2D operators $\O^{s,\pm}_{\Delta,J}(w,\bw)$ defined in~\eqref{eq:2Dop} live on the celestial sphere at the conformal boundary of Minkowski spacetime.
At future null infinity $\mathcal{I}^+$ we use retarded Bondi coordinates $(u,r,z,\bz)$ which are related to the Cartesian coordinates $X^\mu$ by the transformation
\begin{equation}\label{eq:co}
 X^0=u+r\,, \quad X^i=r \hat{X}^i(z,\bz)\,, \quad \hat{X}^i(z,\bz)=\frac{1}{1+z \bar z}(z+\bz,-i(z-\bz),1-z\bz)\,,
\end{equation}
which maps the Minkowski line element to
\begin{equation}\label{eq:gBondi}
ds^2=-du^2-2du dr+2r^2 \gamma_{z \bar z} dz d\bar z \quad \text{with}\quad \gamma_{z \bar z}=\frac{2}{(1+z \bar z)^2}\,.
\end{equation}
One reaches future null infinity by holding $(u,z,\bz)$ fixed and going to large $r$. At $\mathcal{I}^+$ we evaluate the free field mode expansions~\eqref{eq:Aexp}-\eqref{eq:hexp} in a large~$r$ saddle point approximation that identifies the point $(w,\bw)$ on the celestial sphere with the position $(z,\bz)$ towards which a photon or graviton is headed. The $\mathcal{I}^+$ operators creating and annihilating these photons and gravitons are given by
 \begin{equation}\label{eq:hatA}
     \hat{A}_z=-\frac{ie}{\sqrt{2}(2\pi)^2} \int_0^\infty d\omega \left[a_+(\omega,z,\bz)e^{-i\omega (1+z\bz) u}-a^\dagger_-(\omega,z,\bz)e^{i\omega (1+z\bz) u}\right]\,,
 \end{equation}
 and
 \begin{equation}\label{eq:hatH}
  \hat{C}_{zz}=-\frac{i\kappa}{(2\pi)^2} \frac{1}{(1+z\bz)} \int_0^\infty d\omega \left[a_+(\omega,z,\bz)e^{-i\omega (1+z\bz) u}-a^\dagger_-(\omega,z,\bz)e^{i\omega (1+z\bz) u}\right]\,,
 \end{equation}
where $C_{zz}=\lim\limits_{r\to \infty} \frac{1}{r} h_{zz}$.  Meanwhile $\hat{A}_\bz$ and $\hat{C}_{\bz\bz}$ are given by the opposite helicity expressions. 

Starting from the definition of 2D operators~\eqref{eq:2Dop} via the inner products~\eqref{eq:IPspin1}-\eqref{eq:IPspin2} pushed to null infinity and the conformal primary wavefunctions of section~\ref{sec:CPWs}, we can now construct the conformally soft operators at the bottom and top corners of the celestial diamonds that give rise to soft charges in section~\ref{sec:softcharges} and conformal dressings in section~\ref{sec:dressing}.

\subsubsection*{Extrapolate-Style Dictionary}

We close this section by writing out an extrapolate-like dictionary for the external states in celestial CFT.  The dual to celestial CFT correlators is the $\S$-matrix. Equations~\eqref{eq:hatA} and~\eqref{eq:hatH} point to an alternative to the LSZ formalism to extract massless $\S$-matrix elements. Namely, we can prepare the in and out states by integrating the gauge field or metric along a light ray at future or past null infinity.

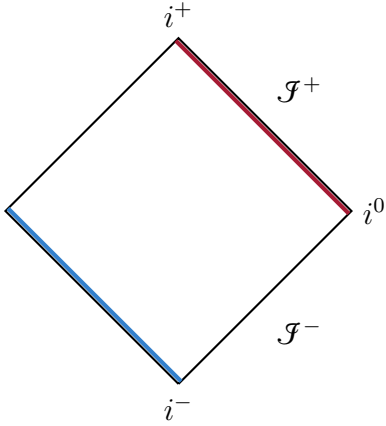
\begin{figure}[th!]
\centering
\begin{tikzpicture}[scale=2.3]
\definecolor{darkgreen}{rgb}{.0, 0.5, .1};
\draw[thick](0,0) --(1,1) node[right] {$i^0$} --(0,2)node[above] {$i^+$} --(-1,1) --(0,0)  node[below] {$i^-$} ;
\draw[ultra thick,dblue] (0+.015,0+.015) -- (-1+.015,1+.015);
\draw[ultra thick,dred] (0-.015,2-.015) -- (1-.015,1-.015);
\node at (1/2+.2,3/2+.2) {$\cal{I}^+$};
\node at (1/2+.2,1/2-.2) {$\cal{I}^-$};
\end{tikzpicture}
\caption{The $\mathcal{O}_{\Delta,J}^\pm(w,\bw)$ correspond to light ray integrals of operator insertions at $\mathcal{I}^\pm$.  For fixed $(w,\bw)$ this is illustrated  by the antipodally placed blue (past) and red (future) contours.
\label{in_out_states}}
\end{figure}

This is illustrated in figure~\ref{in_out_states} (recall the antipodal map across $i^0$~\cite{Strominger:2017zoo}). 
For example, the standard  momentum eigenstate  {\it out} state of a positive helicity helicity photon is prepared via
\be\label{pout}
\langle p,+|=\lim\limits_{r\rightarrow\infty}\int du e^{i\omega(1+z\bz) u}\langle 0|\hat{A}_z
\ee
where $p_\mu=\omega q_\mu(z,\bz)$.  
Let us now define their conformal primary analogs.
Using the following relations \be\scalemath{0.95}{\label{uplus}
\int_{-\infty}^\infty du u_+^{-\Delta} \int_0^\infty d\omega \left[a_\pm e^{-i\omega(1+z\bz) u_+}-a^\dagger_{\mp}e^{i\omega(1+z\bz) u_+}\right]
=\frac{2\pi i^\Delta (1+z\bz)^{\Delta-1}  }{\Gamma(\Delta)} a_{\Delta,\pm s}\,,
}\ee
and
\be\scalemath{0.92}{\label{uminus}
\int_{-\infty}^\infty du u_-^{-\Delta} \int_0^\infty d\omega \left[a_\pm e^{-i\omega(1+z\bz) u_-}-a^\dagger_{\mp}e^{i\omega(1+z\bz) u_-}\right]=\frac{2\pi (-i)^\Delta (1+z\bz)^{\Delta-1} }{\Gamma(\Delta)}  a_{\Delta,\pm s}^\dagger
\,,}
\ee
we see that states of definite conformal dimension are prepared with operators integrated along a light ray of the form
\be\label{lightray}
\mathcal{O}^{1,-}_{\Delta,1}\propto\lim_{r\rightarrow \infty} \int du u_+^{-\Delta} \hat{A}_z(u_+,r,z,\bz), \qquad \mathcal{O}^{2,-}_{\Delta,2}\propto\lim_{r\rightarrow \infty} \frac{1}{r}\int du u_+^{-\Delta} \hat{h}_{zz}(u_+,r,z,\bz) \, ,
\ee
for the $+$ helicity modes, and similarly with the $\bz$ component for the $-$ helicity modes. With an appropriate analytic continuation off the real manifold, we see that light ray operators at future (past) null infinity create the $\emph{out}$ ($\emph{in}$) states from the vacuum. Namely the analog of~\eqref{pout} for a positive helicity outgoing Mellin mode is
 \be
\langle \Delta,z,\bz,+|=\lim_{r\rightarrow \infty} \int du u_+^{-\Delta} \langle 0|\hat{A}_z(u_+,r,z,\bz)\, .
\ee
 Correlation functions of the operators~\eqref{lightray} thus give us $\S$-matrix elements in the conformal basis for any massless theory.

\section{Soft Charges in Celestial Diamonds}
\label{sec:softcharges}

The soft charge corresponding to a given asymptotic symmetry transformation in 4D can be written as an integral over the celestial sphere\footnote{This is following the setup of of~\cite{Banerjee:2018fgd,Banerjee:2019aoy,Banerjee:2019tam} where, starting from the Lorentz representation of the field operators in the free theory, they determine the conformal dimension of the charge. 
}
\begin{equation}\label{eq:Qsoft}
    Q^{soft}_\zeta=\int d^2z\; \zeta(z,\zb)\cdot \O^{soft}(z,\bz)\,,
\end{equation}
where $\zeta$ is the symmetry transformation parameter. We are now going to identify $\O^{soft}$ for the soft charges of asymptotic symmetries in gauge theory and gravity. We will see that they take the form~\cite{Banerjee:2019aoy,Banerjee:2019tam} of primary descendants of the conformally soft radiative fields, i.e. lie at the bottom corners of the corresponding celestial photon and graviton diamonds.\footnote{In gravity, the descendants of the radiative modes that appear in $\O^{soft}$ are related via the constraint equation to the Bondi mass aspect $m_B$ and the the angular momentum aspect $N_z$, which are therefore associated to the bottom corners of the memory diamonds. The values of $\Delta$ we find for the $\O^{soft}$ are consistent with those for the `super-momentum' and `super-angular~momentum' of~\cite{Barnich:2021dta}. }

   \begin{table}[ht!]
   \renewcommand{\arraystretch}{1.3}
    \centering
    \begin{tabular}{r||r||r}
    ${Q^{soft}_\zeta}$&$s=1\hfill$&$s=2\hfill$\\
     \hline\hline
     \rule{0pt}{4ex}leading &${\frac{1}{e^2}}\int d^2z \sqrt{\gamma}~\varepsilon\int du\p_u
     D^B\hat{A}_B$ ~&$ -\frac{2}{\kappa^2}\int d^2z\sqrt{\gamma} f \int du\p_u 
     D^A D^B \hat{C}_{AB}
     $ ~\\[3pt]\hline 
      sub-leading &${-}{\frac{1}{e^2}}\int  d^2z\sqrt{\gamma} \Big[~y^z\int du u \p_u D_z^2\hat{A}^z$~ &${+}\frac{2}{\kappa^2} \int  d^2z\sqrt{\gamma} \Big[~Y^z\int du u \p_u  D_z^3 \hat{C}^{zz}$~~\\
       &$+ y^\bz\int du u \p_u D_\bz^2\hat{A}^\bz\Big]$&$~~~+Y^\bz\int du u \p_u  D_\bz^3 \hat{C}^{\bz\bz}\Big]$
    \end{tabular}
    \caption{
    Soft charge operators for  $\zeta=\{\varepsilon,f, y^A, Y^A\}$ where $A=\{z,\bz\}$. 
    }
    \label{table:qsoft}
\end{table}

Recall the Ward identities for these asymptotic symmetries are equivalent to soft theorems~\cite{Strominger:2017zoo}.\footnote{For the proofs that the momentum space soft theorems give Ward identities for the bulk asymptotic symmetries see~\cite{He:2014cra,He:2014laa,Lysov:2014csa,Kapec:2014opa}.} The goal of~\cite{Donnay:2018neh,Donnay:2020guq} was to show that the charges as defined in~\eqref{eq:2Dop} match ordinary momentum space soft charges summarized in table~\ref{table:qsoft}. We can see this as follows. The soft charges $Q^{soft}_\zeta$ expressed as a celestial sphere integral in~\eqref{eq:Qsoft} correspond to the inner products~\eqref{eq:IPspin1}-\eqref{eq:IPspin2}  with $\Sigma$ pushed to $\mathcal{I}^+$ where $A$ and $h$ are replaced with the field operators~\eqref{eq:hatA} and~\eqref{eq:hatH} while $A'$ and $h'$ are the Goldstone modes of the spontaneously broken asymptotic symmetries. The large-$r$ behavior of these $A^{\Gold/\gold}$ and $h^\Gold$ modes take the form $D_z \varepsilon$ and $D^2_z y^z$ in gauge theory and to $D^2_z f$ and $D^3_z Y^z$ in gravity.  This indeed matches the expressions for the soft charges summarized in table~\ref{table:qsoft}, where we used integration by parts on the celestial sphere to write them in the form~\eqref{eq:Qsoft}.

\subsection{Soft Operators}\label{sec:SoftOp}

The operators $\O^{soft}$ read off by comparing table~\ref{table:qsoft} to~\eqref{eq:Qsoft} involve $u$-integral of the corresponding bulk fields\footnote{  
Using
\begin{equation}\label{bndy}
    \int_{-\infty}^{+\infty} du u^{1-\Delta} \p_u f(u)=(\Delta-1)\int_{-\infty}^{+\infty} du u^{-\Delta} f(u) +\int_{-\infty}^{+\infty} du \p_u \left[u^{1-\Delta} f(u)\right]\,,
\end{equation}
we see that, up to a boundary term, the soft charges can be written in the form~\eqref{lightray}.}.  We will now determine their quantum numbers $(\Delta,J)$ under SL(2,$\mathbb{C}$). To do so we use the inverse Mellin transform~\eqref{eq:inversemellinmode} to express the creation and annihilation operators in~\eqref{eq:hatA} and~\eqref{eq:hatH} in terms of $a^\dagger_{\Delta,\pm s}(z,\bz)$ and $a_{\Delta,\pm s}(z,\bz)$ and note that the $\omega$-integrals take form 
 \begin{equation}\label{Mellinupw}
\int_0^\infty d\omega {\omega}^{1-\Delta} e^{\pm i\omega (1+z\bz) u_\pm} = (\mp i)^{\Delta-2} \Gamma(2-\Delta) u_\pm^{\Delta-2}(1+z\bz)^{\Delta-2} \,,
\end{equation}
where we analytically continue $u\mapsto u_\pm =u\pm i\varepsilon$ to guarantee convergence. 
The $u$-integrals in the soft charges $Q^{soft}_\zeta$ in table~\ref{table:qsoft}
can then be computed\footnote{Note that the $\pm i \varepsilon$ regulator picks out a phase $e^{\pm i \pi}$ when performing the $u$-integrals.} using the generalized distribution~\cite{Donnay:2020guq}
\begin{equation}
    \int_0^\infty du u^{\Delta-1}=2\pi \boldsymbol{\delta}(i\Delta)\,.
\end{equation}
This yields for the annihilation operators
\begin{equation}\label{eq:IDDelta1}
\badat{2}
    & \int_0^\infty d\omega \omega a_\pm(\omega,z,\bz) \int_{-\infty}^{+\infty} du\,u_-^n e^{-i\omega(1+z\bz)u_-}\\
     &\qquad\qquad\qquad\qquad=(-i)^n\pi \lim_{\Delta \to1-n}(\Delta-1+n) a_{\Delta,\pm s}(z,\bz)(1+z\bz)^{-1-n}\,,
\eadat
\end{equation}
and for the creation operators
\begin{equation}\label{eq:IDDelta2}
\badat{2}
    & \int_0^\infty d\omega \omega a^\dagger_\mp(\omega,z,\bz) \int_{-\infty}^{+\infty} du\,u_+^n e^{+i\omega(1+z\bz)u_+}\\
     &\qquad\qquad\qquad\qquad=(+i)^n\pi \lim_{\Delta \to1-n}(\Delta-1+n) a^\dagger_{\Delta,\pm s}(z,\bz)(1+z\bz)^{-1-n}\,.
\eadat
\end{equation}
The leading and sub-leading soft charges correspond to $n=0$ and $n=1$ respectively.
To identify the conformal dimension $\Delta$ and spin $J$ of the operators $\O^{soft}$ we note that the $D_z,D_\bz$ derivatives on the round Bondi sphere turn into the $\p_z,\p_\bz$ derivatives on the flat celestial sphere (which we have been using throughout this paper recalling that the reference direction is now $(z,\bz)$).\footnote{
Recall that for any rank $(m,0)$ tensor $g_{z_1...z_m}$ we have $D_\bz^n g_{z_1...z_m}=\gamma_{z\bz}^n(\gamma^{z\bz}\p_\bz)^n g_{z_1...z_m}$. Meanwhile for any function $g$ we have $\gamma_{z\bz}^n(\gamma^{z\bz}\p_\bz)^n(1+z\bz)^{1-n} g(z,\bz)=(1+z\bz)^{1-n}\p_\bz^n g(z,\bz)$.
}
We thus see that the leading $\O^{soft}$ descend from operators with $\Delta=1$ while the sub-leading $\O^{soft}$ descend from operators which have $\Delta=0$. The dimensions of these descendants are summarized in table~\ref{table:qsoftDJ}.
   \begin{table}[ht!]
    \centering
    \begin{tabular}{l|l|l}
    $(\Delta,J)$&$s=1$&$s=2$\\
     \hline
     \rule{0pt}{4ex}leading &$(2,0)$&$(3,0)$ \\[3pt]
      sub-leading &$(2,\pm 1)$&$(3,\pm 1)$
    \end{tabular}
    \caption{ $(\Delta,J)$ of the soft operators~$\O^{soft}$ defined in table~\ref{table:qsoft}.    }
    \label{table:qsoftDJ}
\end{table}
Using the primary descendant classification of~\cite{Pasterski:2021fjn}, the explicit expressions for $\O^{soft}$ in gauge theory and gravity are as follows.
\paragraph{Leading Soft Photon} The soft operator is a type~II primary descendant operator at level~1 
\begin{equation}
   \O^{soft}_{2,0}=-\frac{1}{\sqrt{2}4\pi e} \lim_{\Delta \to 1}(\Delta-1)  \left[ \partial_\bz (a_{\Delta,+1}+a^\dagger_{\Delta,+1})+\partial_z(a_{\Delta,-1}+a^\dagger_{\Delta,-1})\right]\,.
\end{equation}
We notice the helicity degeneracy of the leading soft photon theorem, with the two terms related by a shadow transform.
\paragraph{Sub-leading Soft Photon}
The type~III primary descendant operator at level~2 
\begin{equation}
     \O^{soft}_{2,+1} =-\frac{i}{4\sqrt{2}\pi e}  \lim_{\Delta\to0}\Delta (1+z\bz)^{-1}\left[\partial^2_z(a_{\Delta,-1}-a^\dagger_{\Delta,-1})\right]\,,
\end{equation}
is the soft operator for the sub-leading conformally soft photon theorem. Similar expressions hold for the opposite 2D spin.
\paragraph{Leading Soft Graviton}
The soft operator is a type~II primary descendant operator at level~2 
\begin{equation}
  \O^{soft}_{3,0}=\frac{1}{4\pi\kappa} \lim_{\Delta \to 1}(\Delta-1)  (1+z\bz)\left[ \partial^2_\bz (a_{\Delta,+2}+a^\dagger_{\Delta,+2})+\partial^2_z(a_{\Delta,-2}+a^\dagger_{\Delta,-2})\right]\,.
\end{equation}
The two terms are related by a shadow transform and display the helicity degeneracy of the leading soft graviton theorem.
\paragraph{Sub-Leading Soft Graviton}
The type~II primary descendant operator 
\begin{equation}
    \O^{soft}_{3,+1}=\frac{i}{4\pi \kappa} \lim_{\Delta\to0}\Delta  \left[\partial^3_z (a_{\Delta,-2}-a^\dagger_{\Delta,-2})\right]\,,
\end{equation}
is the soft operator for the sub-leading conformally soft graviton theorem.
We will see below that it corresponds to a level~3 descendant of the $\Delta=0$ shadow stress tensor~ or a level~1 descendant of the $\Delta=2$ stress tensor. Similar expressions hold for the opposite 2D spin.

\subsection{Memory Diamonds}
We have just seen that the soft charges $Q^{soft}_\zeta$ can be written as a convolution of operators with definite $(\Delta,J)$. From~\cite{He:2015zea,Donnay:2018neh,Himwich:2019dug,Kapec:2016jld} we also know that particular choices of the parameters $\zeta$ in table~\ref{table:qsoft} yield currents in the celestial CFT. 
We will now show how this works for the examples considered in~\cite{Donnay:2018neh} and~\cite{Donnay:2020guq}, which encompass all non-degenerate celestial diamonds for spin-1 and spin-2.  
Without loss of generality we will focus on $J=+s$ radiative primaries, while similar expressions are obtained for $J\mapsto -J$ and $w \mapsto \bar{w}$. 

\subsubsection*{Leading Conformally Soft Photon}
The conformally soft photon current $\J_w$ was defined in~\cite{Donnay:2018neh} as the mode of the bulk field operator~$\hat{A}$ that is extracted from its inner products~\eqref{eq:2Dop} with the spin-1 Goldstone wavefunction
\begin{equation}\label{ConfSoftCurrentJ}
    \J_w={\textstyle \frac{i}{e^2}}(\hat{A},A^\Gold_{1,-1})\,.
\end{equation}
It generates a large U(1) Kac-Moody symmetry
\begin{equation}\label{leadingshift}
    [\J_w,\hat{A}_\mu]=iA^{\Gold}_{1,+1;\mu}=i\nabla_\mu \Lambda_{1,+1}\,,
\end{equation}
where $\Lambda_{1,+1}$ is the potential. At null infinity 
\begin{equation}\label{LambdaScri1} \Lambda_{1,+1}=\frac{1}{\sqrt{2}(z-w)}\equiv \varepsilon_w\,,
\end{equation}
so that we can express the 2D operator~\eqref{ConfSoftCurrentJ} as \be\label{Jw}
\J_w=\int d^2 z\varepsilon_w \O^{soft}_{2,0}\,.
\ee
The current $\J_w$ is associated with the right corner of the spin-1 memory diamond as  shown in the following diagram
\vspace{-1cm}
\begin{equation}
   \raisebox{-1.3cm}{
    \begin{tikzpicture}[scale=0.7]
\definecolor{darkgreen}{rgb}{.0, 0.5, .1};
\draw[thick,->] (-1+.05,1-.05)node[left]{$\J_{\bw}=\p_{\bw}\J$\,} --node[below left]{} (-.05,.05)node[below]{$\beta{\cal O}_{2,0}^{soft}$} ;
\draw[thick,->] (1,1)node[right]{$\J_w=\p_{w}\J$}-- (.05,.05); 
\draw[thick,->] (0,2)node[above]{$\J$}-- (-1+.05,1.05);
\filldraw[black] (0,0) circle (2pt);
\filldraw[black] (-1,1) circle (2pt) ;
\filldraw[black] (1,1) circle (2pt) ;
\filldraw[black] (0,2) circle (2pt) ;
\filldraw[white] (0,0) circle (1pt) ;
\draw[thick,->] (0,2)-- (1-.05,1+.05);
\node[fill=black,regular polygon, regular polygon sides=4,inner sep=1.6pt] at (0,2) {};
\node[fill=white,regular polygon, regular polygon sides=4,inner sep=.8pt] at (0,2) {};
\node at (-3,-1) {};
\node at (3,0) {};
\node at (2,4) {};
\end{tikzpicture}
}
\vspace{-0.2cm}
\end{equation}
where $\beta=-\sqrt{2}\pi$, and the soft operator lies at the bottom corner.  We can further write
 \begin{equation}
\J_w  =\partial_w \J\,,
 \end{equation}
where $\J$ is a conformally soft photon operator at the top of the memory diamond that can be expressed in terms of a $J=0$ pure gauge generalized conformal primary vector with $\Delta=0$
\begin{equation}\label{calJ}
\J={\textstyle \frac{i}{e^2}}(\hat{A},A^{gen,\Gold}_{0,0})\, ,
\end{equation}
where the generalized primary $A^{gen,\Gold}_{0,0}=\p_\bw^{-1}A^\Gold_{1,-1}$ was studied in~\cite{Pasterski:2021fjn}.

\subsubsection*{Leading Conformally Soft Graviton}

The BMS supertranslation current $\P_w=4D^w \N_{ww}$ was defined in~\cite{Donnay:2018neh} as the descendant of the mode of the bulk field operators $\hat{h}$ that is extracted from its inner products~\eqref{eq:2Dop} with the spin-2 Goldstone wavefunction
\begin{equation}\label{ConfSoftCurrentP}
  \N_{ww}={\textstyle -\frac{i}{\kappa^2}} (\hat{h}, h^\Gold_{1,-2})\,.
\end{equation}
The operator $\N_{ww}$ generates a BMS supertranslation symmetry
\begin{equation}\label{leadingshifth}
   [\N_{ww},\hat{h}_{\mu\nu}]=ih^{\Gold}_{1,+2;\mu\nu}=i(\nabla_\mu \xi_{1,+2;\nu}+\nabla_{\nu} \xi_{1,+2;\mu})=i\nabla_\mu \nabla_\nu\Lambda_{1,+2}\,,
\end{equation}
where $\xi_{1,+2;\nu}$ is the supertranslation vector field. At null infinity 
\begin{equation}\label{LambdaScri2}
      \frac{1}{r}\Lambda_{1,+2}=\frac{\bz-\bw}{2(z-w)(1+z\bz)}\equiv -{2}f_{ww}\,,
\end{equation}
so that we can express the 2D operator~\eqref{ConfSoftCurrentP} as
\be\label{Nww}
\N_{ww}=  \int d^2 z f_{ww} \O^{soft}_{3,0}\,.
\ee
The soft operator lies at the bottom of the the leading soft graviton memory diamond, conveniently captured by the following diagram
\begin{equation}
   \raisebox{-2cm}
   {
    \begin{tikzpicture}[scale=.6]
\filldraw[black] (-2,1) circle (2pt)  node [left]{$\N_{\bw\bw}=\frac{1}{2!}\p^2_{\bw}\N$};
\filldraw[black] (2,1) circle (2pt) node [right]{$\N_{w w}=\frac{1}{2!}\p^2_{w}\N$};
\draw[thick] (1+.1414/2,0+.1414/2) arc (45:-135:.1);
\draw[thick] (-1+.1414/2,2+.1414/2) arc (45:-135:.1);
\draw[thick] (0,3) node [above]{$\N$} ;
\draw[->,thick] (0,3) --  (1-.1414/2,2+.1414/2) node [above=4mm]{
};
\draw[->,thick] (1+.1414/2,2-.1414/2) -- (2-.05,1+.05) node [above=4mm]{
};
\draw[->,thick] (0,3) --  (-1+.07,2+.07) node [above=4mm]{
};
\draw[->,thick] (-1-.07,2-.07)-- (-2+.05,1+.05) node [above=4mm]{
};
\draw[->,thick] (1-.07,0-.07)-- (0+.05,-1+.05) ;
\node[fill=black,regular polygon, regular polygon sides=4,inner sep=1.6pt] at (0,3) {};
\node[fill=white,regular polygon, regular polygon sides=4,inner sep=.8pt] at (0,3) {};
\filldraw[black] (0,-1) circle (2pt)  node [below]{ $\beta\mathcal{O}_{3,0}^{soft}$
};
\filldraw[white] (0,-1) circle (1pt);
\draw[->,thick] (2,1) node [below=4mm]{
} --  (1+.07,0+.07) node [below=4mm]{
} ;
\draw[->,thick] (-2,1) node [below=4mm]{
} --  (-1-.1414/2,0+.1414/2) node [below=4mm]{
};
\draw[thick] (-1-.1414/2,0+.1414/2)  arc (135:315:.1);
\draw[->,thick] (-1+.1414/2,0-.1414/2)  -- (0-.05,-1+.05);
\draw[thick] (1-.1414/2,2+.1414/2) arc (135:315:.1);
\end{tikzpicture}
}
\end{equation}
where $\beta=-\frac{\pi}{4}(1+w\bw)^{-1}$. 
Here $\N$ is a conformally soft graviton operator  at the top of the memory diamond, naturally expressed in terms of a $J=0$ pure gauge generalized conformal primary metric with $\Delta=-1$
\begin{equation}
   \N={\textstyle -\frac{i}{\kappa^2}} (\hat{h}, h^{gen,\Gold}_{-1,0})\,,
\end{equation}
where the generalized primary $h^{gen,\Gold}_{-1,0}=2!\p_\bw^{-2}h^\Gold_{1,-2}$ was studied in~\cite{Pasterski:2021fjn}.

\subsubsection*{Sub-leading Conformally Soft Graviton}

In celestial CFT, the stress tensor $\T_{ww}$ is extracted from the inner product~\eqref{eq:2Dop} with the $\Delta=2$ Goldstone wavefunction\cite{Donnay:2020guq}
\begin{equation}\label{Tdef}
 \T_{ww}={\textstyle -\frac{i}{\kappa^2}} (\hat{h},\th^\Gold_{2,-2})\,.
 \end{equation}
 It generates a Virasoro superrotation symmetry
 \begin{equation}
     [\T_{ww},\hat{h}_{\mu\nu}]=i\th^\Gold_{2,+2;\mu\nu}=i\left(\nabla_\mu \xi_{2,+2;\nu}+\nabla_\nu \xi_{2,+2;\mu}\right)\,,
 \end{equation}
 where $\xi_{2,+2}$ is the superrotation vector field. Similarly, the shadow stress tensor $\widetilde{\T}_{\bw\bw}$ is extracted from the inner product~\eqref{eq:2Dop} with the $\Delta=0$ Goldstone wavefunction\cite{Donnay:2020guq}
\begin{equation}\label{SHTdef}
 \widetilde{\T}_{\bw\bw}={\textstyle -\frac{i}{\kappa^2}} (\hat{h},h^\Gold_{0,+2})\,.
 \end{equation}
 It generates a Diff($S^2$) superrotation symmetry
 \begin{equation}
     [\widetilde{\T}_{\bw\bw},\hat{h}_{\mu\nu}]=ih^\Gold_{0,-2;\mu\nu}=i\left(\nabla_\mu \xi_{0,-2;\nu}+\nabla_\nu \xi_{0,-2;\mu}\right)\,,
\end{equation}
where $\xi_{0,-2}$ is the shadow superrotation vector field. In harmonic gauge, diffeomorphisms of the celestial sphere are generated by~\cite{Donnay:2020guq}
\begin{equation}\label{SRdiffeo}
{\textstyle
 \xi_Y=u \alpha \partial_u -\left(\alpha r +u\left(\frac{D^2}{2}+2\right)\alpha \right) \partial_r  + \left(Y^A+\frac{u}{2r} ((D^2+1)Y^A-2D^A\alpha)\right) \partial_A +\dots \,,
 }
\end{equation}
where $Y^A=Y^A(z,\bz)$ is an arbitrary vector field on the sphere and we introduced $\alpha\equiv\frac{1}{2}D_C Y^C$. Since $Y^A$ has two independent components, we expect two linearly independent Ward identities for the two polarizations. As in~\cite{Kapec:2014opa}, we will look at the complexification.
At null infinity, $\xi_{2,+2}$ and $\xi_{0,-2}$ are determined by the vector fields $Y^A_{ww}$ and $Y^A_{\bw\bw}$, respectively, where
\begin{equation}
   Y^z_{ww}=\frac{1}{3!(z-w)}\,, \qquad Y^z_{\bw\bw}=-\frac{(z-w)^2}{2!(\bz-\bw)}\,,
\end{equation}
while $Y^\bz_{ww}=0=Y^\bz_{\bw\bw}$.  The 2D operators~\eqref{Tdef} and~\eqref{SHTdef} can then be expressed as 
\begin{equation}\label{TwwSHTww}
  \T_{ww}=  \int d^2z  Y^z_{ww} \O^{soft}_{3,+1}\,, \qquad
    \widetilde{\T}_{\bw\bw}=  \int d^2z Y^z_{\bw\bw} \O^{soft}_{3,{+}1}\,,
\end{equation}
This soft operator lies at the bottom of the the sub-leading soft graviton memory diamond, conveniently captured by the following diagram
  \begin{equation}
 \raisebox{-2cm}{
\begin{tikzpicture}[scale=0.6]
\definecolor{darkgreen}{rgb}{1, 0,0};
\definecolor{blue}{rgb}{0,0,0};
\filldraw[blue] (-2,2) circle (2pt) ;
\filldraw[blue] (2,0) circle (2pt)  node [right,black]{$\T_{ww}=-\frac{1}{3!}\p^3_{w}\T^w$};
\filldraw[blue] (1,-1) circle (2pt) node [below,black]{$\beta{\cal O}^{soft}_{3,+1}$};
\filldraw[white] (1,-1) circle (1pt) ;
\filldraw[white] (-1,-1) circle (1pt) ;
\draw[thick,->,blue] (2,0) --  (1+.05,-1+.05);
\draw[thick,->,blue] (-1,3)-- (-2+.05,2+.05);
\draw[thick,blue] (0-.1414,0+.1414) arc (135:315:.2);
\draw[thick,blue] (1-.1414,1+.1414) arc (135:315:.2);
\draw[thick,blue] (0-.1414,2+.1414) arc (135:315:.2);
\draw[thick,blue] (-1-.1414,1+.1414) arc (135:315:.2);
\draw[->,thick,blue] (-2,2)  node [left,black]{$\widetilde{\T}_{\bw\bw}=\p_{\bw}\T^w$} --  (-1-.1414,1+.1414);
\draw[->,thick,blue] (-1,3) node [above,black]{$\T^w$} --  (0-.1414,2+.1414);
\draw[->,thick,blue] (-1+.1414,1-.1414) --  (0-.1414,0+.1414);
\draw[->,thick,blue] (0+.1414,0-.1414) -- (1-.05,-1+.05);
\draw[->,thick,blue] (0+.1414,2-.1414) --  (1-.1414,1+.1414);
\draw[->,thick,blue] (1+.1414,1-.1414) -- (2-.05,0+.05);
\node[fill=blue,regular polygon, regular polygon sides=4,inner sep=1.6pt] at (-1,3) {};
\node[fill=white,regular polygon, regular polygon sides=4,inner sep=.8pt] at (-1,3) {};
\end{tikzpicture}
}
 \end{equation}
with $\beta=-\frac{\pi}{3}$ along with its opposite helicity counterpart.  Here we have used~\cite{Pasterski:2021fjn} to write the stress tensor and its shadow as descendants. The operator $\T^w$ is a generalized primary operator with conformal dimension $\Delta=-1$ and spin $J=-1$ that can be defined directly via~\eqref{eq:2Dop} as
 \begin{equation}
\T^w={\textstyle -\frac{i}{\kappa^2}} (\hat{h},h^{gen,\Gold}_{-1,+1})\,,
 \end{equation}
 where the generalized primary $h^{gen,\Gold}_{-1,+1}=-3!\p_\bw^{-3}\th^\Gold_{2,-2}=\p_w^{-1}h^\Gold_{0,+2}$ was studied in~\cite{Pasterski:2021fjn}.

\subsection{Descents and Ascents in the Diamond}

The 2D operators $\J_w$, $\N_{ww}$, $\T_{ww}$ and $\widetilde{\T}_{\bw\bw}$ defined in equations~\eqref{ConfSoftCurrentJ},~\eqref{ConfSoftCurrentP},~\eqref{Tdef} and~\eqref{SHTdef}, respectively, have been shown in~\cite{Donnay:2018neh,Donnay:2020guq} to match the soft charges appearing in table~\ref{table:qsoft}. Moreover, in\cite{Donnay:2020guq} the debate about the asymptotic symmetry groups of Einstein gravity at null infinity involving Diff($S^2$) versus Virasoro was cast into the language of celestial CFT where their generators are related by the 2D shadow transform. This relation is naturally built into celestial diamonds. We will now show that these diamonds also offer a new take on the special symmetry transformation parameters $\{\varepsilon_w,f_{ww},Y^z_{ww},Y^z_{\bw\bw}\}$ 
picked out by the conformal basis. 

Above we discussed how soft charges of spontaneously broken asymptotic symmetries are convolutions of descendants of radiative fields.
The expressions for the leading soft photon~\eqref{Jw} 
\be
\J_w=-\frac{1}{8\pi e}\int d^2 z \frac{1}{(z-w)}\lim\limits_{\Delta\rightarrow 1}(\Delta-1)\Big[ \p_{{\bz}} (a_{\Delta,+1}+a_{\Delta,+1}^\dagger){+} \p_{{z}} (a_{\Delta,-1}+a_{\Delta,-1}^\dagger)\Big]\,,
\ee
the leading soft graviton~\eqref{Nww}
\begin{equation}
\N_{ww}=-\frac{1}{32\pi\kappa}   \int d^2 z \frac{\bz-\bw}{z-w} \lim_{\Delta\to1}(\Delta-1)\Big[ \partial^2_\bz (a_{\Delta,+2}+a^\dagger_{\Delta,+2})+\partial^2_z (a_{\Delta,-2}+a^\dagger_{\Delta,-2})\Big]\,,
\end{equation}
 and the sub-leading soft graviton~\eqref{TwwSHTww}
\be
\badat{2}\label{GreensSubgrav}
\T_{ww}&=\frac{i}{24\pi \kappa} \int d^2 z \frac{1}{z-w} \lim_{\Delta\to0}\Delta\Big[ \partial^3_z (a_{\Delta,-2}-a^\dagger_{\Delta,-2})\Big]\,,\\
\widetilde{\T}_{\bw\bw}&=-\frac{i}{8\pi \kappa}   \int d^2 z \frac{(z-w)^2}{\bz-\bw} \lim_{\Delta\to0}\Delta\Big[ \partial^3_z (a_{\Delta,-2}-a^\dagger_{\Delta,-2})\Big]\,,
\eadat
\ee
hint that special convolutions take us back up to radiative primary operators. This gives us an interesting interpretation for the parameters $\{\varepsilon_{w},f_{ww},Y^z_{ww},Y^z_{\bw\bw}\}$: they are Green's functions that invert the descendants appearing in the soft charges. This can be seen from\footnote{
This follows from the relation
\begin{equation}
\label{shadow_Ia_Ib}
 \p_{\bw'}^{\bar \n} \frac{(\bw'-\bw)^{\bar{\n}-1}}{ (w'-w)^{\n+1}}
   =2\pi (\bar{\n}-1)! \frac{(-1)^{\n}}{\n!} \p_{w'}^\n\delta^{(2)}(w'-w) \,,
\end{equation}
with $\p_z \frac{1}{\bar{z}}=2\pi \delta^{(2)}(z)$ and for $\n=0$, and similarly for its complex conjugate, which was used in~\cite{Pasterski:2021fjn} to prove the shadow relation in the celestial diamond.}
\be\label{greensfn}
\p_\bw^{-\bk}\Phi(w,\bw)=\frac{(-1)^\bk}{2\pi(\bk-1)!}\int d^2 w' \frac{(\bw-\bw')^{\bk-1}}{w-w'}\Phi(w',\bw')
\ee
and similarly for $\p_w^{-k}$. 
The meromorphic vector field $Y^z_{ww}$ and the special Diff($S^2$) vector field $Y^z_{\bw\bw}$ of~\cite{Donnay:2020guq,Campiglia:2014yka} invert the descendant in the definition of the sub-leading soft charge, i.e. lift the bottom corner of the diamond up to the stress tensor and its shadow, respectively. Namely, we have 
\be\label{zrel}
Y^z_{ww}\Rightarrow \p_\bz^{-1},~~~Y^z_{\bw\bw}\Rightarrow \p_z^{-3}\,,
\ee
and by symmetry in $z\leftrightarrow w$, we also have
\be\label{wrel}
Y^z_{ww}\Rightarrow \p_\bw^{-1},~~~Y^z_{\bw\bw}\Rightarrow \p_w^{-3}\,.
\ee
The relation~\eqref{zrel} is the one relevant to the superrotation charge~\eqref{GreensSubgrav}, while the descendancy relations~\eqref{wrel} ensure consistency between our expressions for $\T_{ww}$ and $\widetilde{\T}_{\bw\bw}$ in terms of $\T^w$. Similar Green's functions appear for the other soft theorems. As seen in section~\ref{sec:SoftOp}, the special symmetry parameters $\{\varepsilon_{w},f_{ww},Y^z_{ww},Y^z_{\bw\bw}\}$ have precisely the right factors of $(1+z\bz)$ to guarantee that this integral kernel matches the appropriate uplifting Green's function.

Note that the memory effects~\cite{Strominger:2014pwa,Pasterski:2015tva} are defined up to a kernel for the global symmetries~\cite{Pasterski:2021fjn}.  This describes an ambiguity in the corresponding Green's functions lifting us from the bottom to the left and right corners of the celestial diamonds.  This ambiguity also applies when the corresponding Green's functions are used to go from the left and right corners to the top of the diamond.  
We leave to future work the question of whether or not to include the modes \{$\J$,$\N$,$\T^A$\}, appearing at the top of the memory diamonds, in the phase space.\footnote{When we formally invert the descendancy relations with these Green's functions, we smear any contact term violations or sources.  This explains why the gauge fixings can be expected to be more relaxed for the top corners. 
} We will turn to their analogs in the Goldstone diamonds in the next section. 

\subsection{Generalized Celestial Currents}
\label{sec:gen_j}

Now that we have discussed the corners of the diamond, let us point out that certain operators which belong to the edges are also relevant to the celestial CFT literature. As mentioned above, within $S$-matrix elements the soft operators at the bottom of the memory diamonds reduce to contact terms. This arises from the fact that memory effects are determined by the constraint equations and the matter distributions in perturbative ${S}$-matrix elements are localized to isolated points on the celestial sphere. The associated soft charges were expressed above as integrals over the full celestial sphere. However, the original motivation for celestial CFT arose from an isomorphism between Ward identities for the asymptotic symmetries in 4D and ones for holomorphic currents in 2D, and so one would like to use codimension~one charges to describe the symmetry enhancements. Indeed, this would seem essential to any applications of radial quantization techniques to celestial CFTs.

We have already encountered bona fide holomorphic currents for the leading soft gauge boson~\cite{He:2015zea} and shadow transformed sub-leading soft graviton~\cite{Kapec:2014opa,Kapec:2016jld}, namely the soft photon current $\J_w$ and the stress tensor $\T_{ww}$. In addition, two instances of exotic `currents' have been proposed: the supertranslation current~\cite{Strominger:2013jfa} associated to the leading soft graviton and a sub-leading soft photon current~\cite{Himwich:2019dug}.
In our language these are just descendants of the primary operators at the left and right corners of the diamonds. Indeed in~\cite{Donnay:2018neh} it was observed that the supertranslation current can be written as the level-1 descendant $\P_w=8\p_\bw \N_{ww}$ of the operator associated to the $\Delta=1$ radiative spin-2 primary. Similarly, the sub-leading soft photon current can be expressed as the level-1 descendant of the operator associated to the $\Delta=0$ radiative spin-1 primary; this descendent thus lies halfway between the radiative $\Delta=0$ primary and its type~III primary descendant given by the radiative $\Delta=2$ shadow primary.

The list of such celestial currents does not stop there. In fact, for any spin-$s$ celestial memory diamond we can define the associated celestial current as the level-1 `ascendant' of the soft charge operator, i.e. currents $j$ and $\bar j$ such that
\begin{equation}\label{jjbar}
  \p_\bw  j=\O^{soft}_{\Delta,J}\quad \text{and} \quad  \p_w \bar{j}=\O^{soft}_{\Delta,J}\,.
\end{equation}
Note that the data in these `currents' is of course already contained in their primary parents at the left or right corners of the diamonds. However $j$ and $\bar{j}$ have the nice property of satisfying a canonical `first order conservation equations' instead of the `higher derivative conservation equations'\footnote{These already appeared in several contexts, see e.g.  \cite{Dolan:2001ih, Brust:2016gjy} and references therein. } of the left and right corner operators. The first is holomorphic away from sources, the second anti-holomorphic.  We can then define Laurent mode operators 
 \be
Q_n= \frac{1}{2\pi i}\oint dw w^{\Delta-1+n}j(w),~~~j(w)=\sum_{n\in \mathbb{Z}}w^{-n-\Delta}Q_n.
 \ee
 Because the soft operators are contact terms in celestial amplitudes, we expect that the $Q_n$ can be interpreted as charges and that the hard operators are charged.
 
 From figure~9 of~\cite{Pasterski:2021fjn}\footnote{We note that for the finite dimensional multiplets discussed in section~4.1 of~\cite{Pasterski:2021fjn}, there are a series of \mbox{(anti-)}holomorphic operators, corresponding to the states in the multiplet with largest $\Delta$ for fixed $J$.  As illustrated in figure~4 of that reference, these each have a level-1 descendant 
which vanishes. Since the descendants vanish, the corresponding Ward identities would have no source terms.}  it follows that these operators have scaling dimension $\Delta=s$.  Moreover there are $2s$ such $j$'s and $2s$ such $\bar{j}$'s. For example, the sub-leading soft graviton diamond tells us that the stress tensor $\T_{ww}=-\frac{\pi}{3}\partial_\bw^{-1}\O^{soft}_{3,+1}$ is joined by the $\Delta=2$ celestial current $\pi\partial_w^{-1}\O^{soft}_{3,+1}=\frac{1}{2!}\p_w^2 \widetilde{\T}_{\bw\bw}$ which is a descendant but not a primary, similarly to the supertranslation current. We get the following OPE for the Mellin operators of~\cite{Fotopoulos:2019vac}
 \be
 \lim_{\Delta\rightarrow 0}\Delta \p_w^2 \O_{\Delta,-2}(w,\bw) \O_{\Delta_j,J_j}(w_j,\bw_j)\sim \frac{2}{\bw-\bw_j}\p_{w_j} \O_{\Delta_j,J_j}\,.
\ee

\section{Conformal Dressings in Celestial Diamonds}
\label{sec:dressing}

Conformal Faddeev-Kulish dressings which render amplitudes infrared finite were found in~\cite{Arkani-Hamed:2020gyp} to be given by Goldstone boson insertions. As we will show, the latter naturally arise from the generalized conformal primaries forming the top corners of our celestial diamonds. While~\cite{Arkani-Hamed:2020gyp} focused on dressings related to the leading conformally soft photon and graviton theorems, here we identify the sub-leading conformal Faddeev-Kulish dressings in gauge theory and gravity.  

\subsection{Conformally Soft Modes}

\subsubsection*{Leading conformally soft modes}

The conformally soft photon and BMS supertranslation currents generate but are invariant under large gauge transformations. We can also define 2D operators that shift. These Goldstone currents are extracted from the inner product of the bulk field operators $\hat{A}$ and $\hat{h}$ with the conformally soft photon and graviton wavefunctions\footnote{With our choice of normalization for the conformal primaries~\eqref{Aconfsoft}-\eqref{hconfsoft} and~\eqref{AhCS}-\eqref{eq:phiCS} we have $i(A^{\CS'},A^\Gold)=i(h^{\CS'},h^\Gold)=(2\pi)^2\delta^{(2)}(w-w')$. 
}~\cite{Donnay:2018neh,Arkani-Hamed:2020gyp} 
\begin{equation}\label{GoldtoneCurrents}
 \S_w ={\textstyle -\frac{i}{2\sqrt{2}\pi}} (\hat{A},A^{\CS'}_{1,-1})\,, \quad  \C_{ww} ={\textstyle -\frac{i}{2\pi}} (\hat{h},h^{\CS'}_{1,-2})\,,
\end{equation}
and analogous expressions for $\S_\bw$ and $\C_{\bw\bw}$. These 2D operators are associated with the left and right corners of the celestial Goldstone diamonds for the leading soft photon and graviton:
\vspace{-0.5cm}
\begin{equation}
   \raisebox{-1.55cm}{
   \begin{tikzpicture}[scale=0.7]
\definecolor{darkgreen}{rgb}{.0, 0.5, .1};
\draw[thick,->] (-1+.05,1-.05)node[left]{$\S_{\bw} = \partial_\bw \S$\,} --node[below left]{} (-.05,.05)node[below]{$\p_w\p_\bw \S$} ;
\draw[thick,->] (1,1)node[right]{$\S_w=\partial_w \S$}-- (.05,.05); 
\draw[thick,->] (0,2)node[above]{$\S$}-- (-1+.05,1.05);
\filldraw[black] (0,0) circle (2pt);
\filldraw[black] (-1,1) circle (2pt) ;
\filldraw[black] (1,1) circle (2pt) ;
\filldraw[black] (0,2) circle (2pt) ;
\filldraw[white] (0,0) circle (1pt) ;
\draw[thick,->] (0,2)-- (1-.05,1+.05);
\node[fill=black,regular polygon, regular polygon sides=4,inner sep=1.6pt] at (0,2) {};
\node[fill=white,regular polygon, regular polygon sides=4,inner sep=.8pt] at (0,2) {};
\node at (-3,-1) {};
\node at (3,0) {};
\node at (2,4) {};
\end{tikzpicture}
}
\qquad
 \raisebox{-2cm}
   {
    \begin{tikzpicture}[scale=.6]
\filldraw[black] (-2,1) circle (2pt)  node [left]{$\C_{\bw\bw}=\frac{1}{2!}\p^2_{\wb} \C$};
\filldraw[black] (2,1) circle (2pt) node [right]{$\C_{w w}=\frac{1}{2!}\p^2_{w} \C$};
\draw[thick] (1+.1414/2,0+.1414/2) arc (45:-135:.1);
\draw[thick] (-1+.1414/2,2+.1414/2) arc (45:-135:.1);
\draw[thick] (0,3) node [above]{$\C$} ;
\draw[->,thick] (0,3) --  (1-.1414/2,2+.1414/2) node [above=4mm]{
};
\draw[->,thick] (1+.1414/2,2-.1414/2) -- (2-.05,1+.05) node [above=4mm]{
};
\draw[->,thick] (0,3) --  (-1+.07,2+.07) node [above=4mm]{
};
\draw[->,thick] (-1-.07,2-.07)-- (-2+.05,1+.05) node [above=4mm]{
};
\draw[->,thick] (1-.07,0-.07)-- (0+.05,-1+.05) ;
\node[fill=black,regular polygon, regular polygon sides=4,inner sep=1.6pt] at (0,3) {};
\node[fill=white,regular polygon, regular polygon sides=4,inner sep=.8pt] at (0,3) {};
\filldraw[black] (0,-1) circle (2pt)  node [below]{ $\frac{1}{2!}\p^2_{w} \frac{1}{2!}\p^2_{\wb} \C$
};
\filldraw[white] (0,-1) circle (1pt);
\draw[->,thick] (2,1) node [below=4mm]{
} --  (1+.07,0+.07) node [below=4mm]{
} ;
\draw[->,thick] (-2,1) node [below=4mm]{
} --  (-1-.1414/2,0+.1414/2) node [below=4mm]{
};
\draw[thick] (-1-.1414/2,0+.1414/2)  arc (135:315:.1);
\draw[->,thick] (-1+.1414/2,0-.1414/2)  -- (0-.05,-1+.05);
\draw[thick] (1-.1414/2,2+.1414/2) arc (135:315:.1);
\end{tikzpicture}
}
\vspace{-0.2cm}
\end{equation}
 The top corners of these diamonds are associated to the Goldstone bosons $\S$ and $\C$ which are obtained from an inner product with the non-gauge $J=0$ generalized primary vector and metric from which the $\CS'$ wavefunctions of spin-1 and spin-2 descend: 
\begin{equation}\label{SCgenprim}
    \S={\textstyle -\frac{i}{2\sqrt{2}\pi}}(\hat{A},A^{gen,\CS'}_{0,0})\,, \quad \C={\textstyle -\frac{i}{2\pi}}(\hat{h},h^{gen,\CS'}_{-1,0})\,,
\end{equation}
 where $A^{gen,\CS'}_{0,0}=\p_\bw^{-1}A^{\CS'}_{1,-1}$ and $h^{gen,\CS'}_{-1,0}=2!\p_\bw^{-2}h^{\CS'}_{1,-2}$ were defined in~\cite{Pasterski:2021fjn}.
The Goldstone currents~\eqref{GoldtoneCurrents} are their type~I primary descendants 
as illustrated in the diagram above.

The Goldstone bosons $\S$ and $\C$ turn out to select conformal dressings for amplitudes~\cite{Arkani-Hamed:2020gyp} as we will review below. We can thus understand conformal dressings as arising from the top corners of the celestial photon and graviton diamonds.

\subsubsection*{Sub-leading conformally soft modes}

Conformally soft theorems exist beyond leading order and we can use the corresponding $\Memo$ and $\memo$ wavefunctions to define 2D operators analogous to the Goldstone currents~\eqref{GoldtoneCurrents} which are relevant for sub-leading conformal dressings. 

Indeed, besides the spin-2 wavefunction with $\Delta=2$ which corresponds to (Virasoro) superrotations and gives rise to the celestial stress tensor, there is another spin-2 conformal primary wavefunctions with $\Delta=2$ that is not pure gauge. In~\cite{Ball:2019atb} its inner product with the bulk field operator $\hat h$ was defined as a `dual stress tensor'  
 \begin{equation}\label{Tdualdef}
   \Y_{ww}={\textstyle -\frac{i}{2\pi}}(\hat{h},h^\Memo_{2,-2})\,,
 \end{equation}
 with an analogous expression for $\Y_{\bw\bw}$. These 2D operators are associated with the left and right corners of the Goldstone diamonds associated to the sub-leading soft graviton: 
  \begin{equation}
   \raisebox{-2cm}{
\begin{tikzpicture}[scale=0.6]
\definecolor{darkgreen}{rgb}{1, 0,0};
\definecolor{blue}{rgb}{0,0,0};
\filldraw[blue] (-2,2) circle (2pt) ;
\filldraw[blue] (2,0) circle (2pt)  node [right,black]{$\Y_{ww}=\frac{1}{3!}\p^3_{w}\Y^{w}$};
\filldraw[blue] (1,-1) circle (2pt) node [below,black]{$-\frac{1}{3!}\p^3_{w}\p_\bw \Y^w$};
\filldraw[white] (1,-1) circle (1pt) ;
\filldraw[white] (-1,-1) circle (1pt) ;
\draw[thick,->,blue] (2,0)-- (1+.05,-1+.05);
\draw[thick,->,blue] (-1,3)-- (-2+.05,2+.05);
\draw[thick,blue] (0-.1414,0+.1414) arc (135:315:.2);
\draw[thick,blue] (1-.1414,1+.1414) arc (135:315:.2);
\draw[thick,blue] (0-.1414,2+.1414) arc (135:315:.2);
\draw[thick,blue] (-1-.1414,1+.1414) arc (135:315:.2);
\draw[->,thick,blue] (-2,2)  node [left,black]{$-\p_\bw \Y^w$} --  (-1-.1414,1+.1414);
\draw[->,thick,blue] (-1,3) node [above,black]{$\Y^w$} --  (0-.1414,2+.1414);
\draw[->,thick,blue] (-1+.1414,1-.1414) --  (0-.1414,0+.1414);
\draw[->,thick,blue] (0+.1414,0-.1414) -- (1-.05,-1+.05);
\draw[->,thick,blue] (0+.1414,2-.1414) --  (1-.1414,1+.1414);
\draw[->,thick,blue] (1+.1414,1-.1414) -- (2-.05,0+.05);
\node[fill=blue,regular polygon, regular polygon sides=4,inner sep=1.6pt] at (-1,3) {};
\node[fill=white,regular polygon, regular polygon sides=4,inner sep=.8pt] at (-1,3) {};
\end{tikzpicture}
}
 \end{equation}
The top corners of the diamond containing~\eqref{Tdualdef} is associated with the superrotation Goldstone mode operator~\cite{Himwich:2019qmj} ${\Y}^w$ obtained from the $J=-1$ generalized primary metric with $\Delta=-1$, namely
 \begin{equation}\label{Fw}
     {\Y}^w={\textstyle -\frac{i}{2\pi}}(\hat{h},h^{gen,\Memo}_{-1,+1})\,,
 \end{equation}
  where $h^{gen,\Memo}_{-1,+1}=3!\p_\bw^{-3}h^\Memo_{2,-2}$ was defined in~\cite{Pasterski:2021fjn}. Meanwhile the other diamond is associated with ${\Y}^\bw$. The dual stress tensor~\eqref{Tdualdef} is a type~I primary descendant of~\eqref{Fw} as shown in the diagram.
  
We can consider analogous statements for the sub-leading soft theorem in gauge theory arising from the spin-1 wavefunctions with $\Delta=2$. The electromagnetic analogue of~\eqref{Tdualdef} is the 2D operator
\begin{equation}\label{Rw}
    \y_w ={\textstyle -\frac{i}{2\sqrt{2}\pi}}(\hat{A},A^{\memo}_{2,-1})\,,
\end{equation}
and an analogous expression for $\y_\bw$. Since the celestial diamonds associated to the  sub-leading soft photon are degenerate with zero area all corners are described by radiative spin-1 conformal primary wavefunctions:
\begin{equation}
\raisebox{-1.2cm}{
    \begin{tikzpicture}[scale=.7]
\filldraw[black] (1,0)node[right]{$ \y_w $} circle (2pt) ;
\filldraw[black] (-1,2)node[left]{$\y^w $} circle (2pt) ;
\draw[thick] (0-.1414,1+.1414) arc (135:315:.2);
\draw[->,thick] (-1,2) --  (0-.1414,1+.1414);
\draw[->,thick] (0+.1414,1-.1414) -- (1-.03,0.03);
\end{tikzpicture}
} 
\end{equation}
The above $\Delta=2$ primary descends from a $\Delta=0$ primary, which is also its shadow transform. This parent corresponds to the 2D operator
\begin{equation}\label{Rwtop}
    \y^w ={\textstyle -\frac{i}{2\sqrt{2}\pi}}(\hat{A},\tA^{\memo}_{0,+1})\,.
\end{equation}
Again, an analogous expression exists for $\y^\bw$. We thus see that all of these operators are already in the spectrum of the theory and~\eqref{Rw} is the type~III primary descendant of~\eqref{Rwtop} with
 \begin{equation}
       \y_w  =-\frac{1}{2!}\partial_w^2 \y^w \,,
 \end{equation}
 as shown in the diagram.

In the following we will show that ${\Y}^w$ and $\y^w$ select sub-leading conformal dressings for amplitudes which thus assigns physical significance to the top corners of celestial Goldstone diamonds.

\subsection{Conformally Soft Dressings} 

The Faddeev-Kulish dressings for QED and gravity were constructed in~\cite{Choi:2019rlz} up to sub-leading order in the soft expansion and to leading order in the coupling constants $e$ and $\kappa$ as linearized coherent states that respect charge conservation. For a single (charged) particle $j$ we have the QED dressing\footnote{The leading terms arise from the eikonal limit and exponentiate the soft exchanges. Meanwhile, the sub-leading terms were written suggestively in the exponential in~\cite{Choi:2019rlz}, as~\eqref{EMdressing} and~\eqref{gravitydressing}.  These sub-leading terms are valid to leading order in the couplings. 
} 
 \be\label{EMdressing}
W_{e,j}[\phi_e]=\exp\Big\{-e\int \frac{d^3 k}{(2\pi)^3}\frac{\phi_e(\vec{k})}{2k^0}\frac{Q_j}{p_j\cdot k}\left[\left(p_j^\mu -ik_\nu J_j^{\nu\mu}\right)a_\mu-\left(p_j^\mu +ik_\nu J_j^{\nu\mu}\right)a_\mu^\dagger\right]\Big\}\,,
\ee
and the gravitational dressing
\be\label{gravitydressing}
W_{G,j}[\phi_G]=\exp\Big\{-\frac{\kappa}{2}\int \frac{d^3 k}{(2\pi)^3}\frac{\phi_g(\vec{k})}{2k^0}\frac{p_j^\mu}{p_j\cdot k}\left[\left(p_j^\nu -ik_\rho J_j^{\rho\nu}\right)a_{\mu\nu}-\left(p_j^\nu +ik_\rho J_j^{\rho\nu}\right)a_{\mu\nu}^\dagger\right]\Big\}\,,
\ee
where $\phi_e(\vec{k})$ and $\phi_G(\vec{k})$ are arbitrary functions obeying $\phi_e(0)=1=\phi_G(0)$.
Here $Q_j$ is the charge of the $j$-th particle while 
$p_j^\mu$ and $J_j^{\mu\nu}=i(p^\mu_j \partial_{p_{j\nu}}-p^\nu_j \partial_{p_{j\mu}})$ are its momentum and angular momentum which we parametrize such that $p_j\cdot k=-2 \omega \eta_j \omega_j|w-w_j|^2$ where $\eta_j=\pm1$ for outgoing (incoming) particles. The dressed asymptotic states for QED and gravity are then given by
\begin{equation}
    W_{e,j}[\phi_e]|p_j,Q_j\rangle\,, \quad W_{G,j}[\phi_G]|p_j\rangle\,.
\end{equation}
The special choice of Faddeev-Kulish dressing $\phi_e(\vec{k})=1=\phi_G(\vec{k})$ was shown in~\cite{Arkani-Hamed:2020gyp} to yield the exponentiated Goldstone bosons $\S$ and $\C$ as the leading conformally soft dressings. This prescription will furthermore give us the sub-leading conformally soft dressings in gauge theory and gravity in terms of the operators $\Y^w$ and $\y^w$. 

To see this we make use of the fact that the integrands in~\eqref{EMdressing} and~\eqref{gravitydressing} have definite powers of $\omega$ which, as in section~\ref{sec:softcharges}, turns the creation and annihilation operators in the momentum basis via the inverse Mellin transform~\eqref{eq:inversemellinmode} into operators with definite conformal dimension. 
The contributions to the leading conformally soft dressings are obtained from
\begin{equation}\label{leading1}
  [\partial_w \log(p_j \cdot q)]  Q_j  = Q_j\frac{1}{w-w_j}\,,
\end{equation}
and
\begin{equation}\label{leading2}
     [\partial_w \log(p_j \cdot q)]^2 (p_j \cdot q)
     =-2\eta_j \omega_j \frac{\bw-\bw_j}{w-w_j}\,.
\end{equation}
For the sub-leading conformally soft dressing we note that the angular momentum $J^{\mu\nu}_j$ contracted with the null vector $q_\mu$ and the polarization vector $\sqrt{2}\epsilon_{w;\nu}=\partial_w q_\nu$ can be expressed in terms of the generators of rotations $\vec{J}^{(j)}$ and boosts $\vec{K}^{(j)}$ 
\be
(iq_\rho \p_w q_\nu J^{\rho\nu}_j)=-(1-\bw^2)(J_1^{(j)}+iK_1^{(j)})+i(1+\bw^2)(J_2^{(j)}+iK_2^{(j)})+2\bw (J_3^{(j)}+iK_3^{(j)})\,.
\ee
This can be evaluated in terms of a simple action on the reference direction $\{w_j,\bw_j\}$ using the results in appendix~B of~\cite{Pasterski:2021fjn}. We have the following relations between the SL(2,$\mathbb{C}$) Lorentz generators $\vec{J}$ and $\vec{K}$ and celestial derivatives $\partial_{w}$ and $\partial_{\bw}$ (momentarily dropping the $j$ label)
\begin{equation}
\badat{3}
\left(J_3-i K_3-2w\p_w -2h\right)\Phi_{\Delta,J}&=0\,,\\
\left(-J_1+i J_2+i K_1+K_2-2\p_w \right)\Phi_{\Delta,J}&=0\,,\\
\left(-i K_1+ K_2+J_1+i J_2-2w^2\p_w- 4h w\right)\Phi_{\Delta,J}&=0\,,
\eadat
\ee
and similar expressions for the barred quantities obtained by taking the complex conjugates of the differential operators. 
Using these relations we find 
\begin{equation}\label{sub-leading1}
 (p_j \cdot q)^{-1} Q_j   (iq_\nu \p_w q_\mu J_j^{\nu\mu})
 = \frac{Q_j}{\eta_j\omega_j}  \frac{1}{w-w_j}\left[(\bw-\bw_j)\partial_{\bw_j}-2\bh_j\right]\,,
\end{equation}
and
\begin{equation}\label{sub-leading2}
 [\partial_w \log(p_j \cdot q)]   (iq_\rho \p_w q_\nu J_j^{\rho\nu})= -2\frac{\bw-\bw_j}{w-w_j}\left[(\bw-\bw_j)\partial_{\bw_j}-2\bh_j\right]\,.
\end{equation}
From~\eqref{leading1}-\eqref{leading2} and~\eqref{sub-leading1}-\eqref{sub-leading2}, and similar expressions for the opposite helicity terms, we can now infer the conformal Faddeev-Kulish dressings. 

The conformally soft QED dressing up to sub-leading order in the soft expansion and to leading order in $e$ is given by \begin{equation}\label{Wefinal}\scalemath{0.98}{
W_{e,j}=\exp\left\{-iQ_j\S(w_j,\bw_j)+\frac{Q_j}{\eta_j\omega_j} \left(2h_j\p_{w_j} \y^{w_j}+\y^{w_j}\p_{w_j} +2\bh_j\p_{\bw_j} \y^{\bw_j}+\y^{\bw_j}\p_{\bw_j} \right)  \right\}\,},
\end{equation}
where the leading dressing arises from the Goldstone boson 
\begin{equation}\label{Sdressing}
\S(w_j,\bw_j)={\frac{ie}{4\sqrt{2}\pi^2}}\int_0^\infty d\omega\Big[ \p_{\bw_j}^{-1}\left( a_{-}-a_{+}^\dagger\right) +\p_{w_j}^{-1}\left( a_{+}-a_{-}^\dagger\right) \Big]\,,
\end{equation}
while the operators responsible for the sub-leading dressing are given by 
\begin{equation}\label{ydressing}
\badat{2}
\y^{w_j}&={\frac{e}{4\sqrt{2}\pi^2}}\int_0^\infty d\omega \omega \Big[ \p_{w_j}^{-2}\left( a_{+}+a_{-}^\dagger\right) \Big]\,, \\ \y^{\bw_j}&={\frac{e}{4\sqrt{2}\pi^2}}\int_0^\infty d\omega \omega \Big[ \p_{\bw_j}^{-2}\left( a_{-}+a_{+}^\dagger\right) \Big]\,.
\eadat
\end{equation}
Comparing the $\omega$-integrals to the Mellin transform~\eqref{mellinmode} we see that the operators in~\eqref{Sdressing} and~\eqref{ydressing} may be interpreted as a smearing of the conformally soft radiative modes of conformal dimensions $\Delta=1$ and $\Delta=2$, respectively. The first term in~\eqref{Wefinal}, given by the Goldstone boson~\eqref{Sdressing}, matches~\cite{Arkani-Hamed:2020gyp}. With the remaining terms in~\eqref{Wefinal} given by the operators~\eqref{ydressing} we have extended conformally soft Faddeev-Kulish dressing of~\cite{Arkani-Hamed:2020gyp} to the sub-leading case. 

For gravity we find the conformally soft dressing up to sub-leading order in the soft expansion and to leading order in $\kappa$ to be given by 
\begin{equation}\label{Wgfinal}
W_{g,j}
=\exp\left\{-{i \eta_j\omega_j}\C(w_j,\bw_j) +2\left(h_j\p_{w_j} {\Y}^{w_j}+{\Y}^{w_j}\p_{w_j}
 +\bh_j\p_{\bw_j} {\Y}^{\bw_j} +{\Y}^{\bw_j}\p_{\bw_j} \right) \right\}\,,
\end{equation}
where the leading dressing arises from the Goldstone boson 
\begin{equation}\label{Cdressing} 
\C(w_j,\bw_j)={\frac{i\kappa}{8\pi^2}}\int_0^\infty d\omega\Big[ \p_{\bw_j}^{-2}\left( a_{-}-a_{+}^\dagger\right) +\p_{w_j}^{-2}\left( a_{+}-a_{-}^\dagger\right) \Big]\,,
\end{equation}
while the operators responsible for the sub-leading dressing are given by
\begin{equation}\label{Ydressing}
\badat{2}
\Y^{w_j}&={\frac{\kappa}{8\pi^2}}\int_0^\infty d\omega \omega \Big[ \p_{w_j}^{-3}\left( a_{+}+a_{-}^\dagger\right) \Big]\,, \\
\Y^{\bw_j}&={\frac{\kappa}{8\pi^2}}\int_0^\infty d\omega \omega \Big[ \p_{\bw_j}^{-3}\left( a_{-}+a_{+}^\dagger\right) \Big]\,.
\eadat
\end{equation}
Comparing to~\eqref{mellinmode} we see that the operators in~\eqref{Cdressing} and~\eqref{Ydressing} are a smearing of the conformally soft radiative modes of conformal dimension $\Delta=1$ and $\Delta=2$, respectively. The first term in~\eqref{Wgfinal}, given by the Goldstone boson~\eqref{Cdressing}, matches~\cite{Arkani-Hamed:2020gyp} while the remaining terms in~\eqref{Wgfinal} given by the operators~\eqref{Ydressing} extend their result to the sub-leading case. 

We have written the Green's functions in a compact form in order to emphasize the fact that the above expressions are local in terms of the dressing fields. As in~\cite{Cheung:2016iub,Arkani-Hamed:2020gyp,Haehl:2019eae} one can write these in terms of the integral kernel~\eqref{greensfn}. 
Moreover, we can replace any appearance of $\omega_j$ in the dressings by the appropriate powers of the operator $\hat{\omega}_j$, whose action in the conformal basis is to shift the conformal dimension $\Delta_j$ of a primary operator\footnote{The operator $\widehat{\omega}$ is related to a particular combination of the components of the translation generator $P^\mu=q^\mu e^{\partial/\partial_\Delta}$, given by ${\textstyle \frac{1}{2}}(P^0+P^3)=e^{\partial/\partial_\Delta}$ in~\cite{Stieberger:2018onx} and denoted by $P$ in~\cite{Arkani-Hamed:2020gyp}. Namely, $P=\eta \hat{\omega}$ where $\eta=\pm1$ for outgoing (incoming) particles.}
\be\label{omegahat} 
 \widehat{\omega}_j \O(p_j)=\omega_j \O(p_j)~~\Rightarrow~~ \widehat{\omega}_j \O_{\Delta_j}(w_j,\bw_j)=\O_{\Delta_j+1}(w_j,\bw_j)\,.
\ee
From~\eqref{Wefinal} and~\eqref{Wgfinal} we thus see that the sub-leading QED dressing incorporates a shift by $\Delta_j\to \Delta_j-1$ while the leading gravitational dressing shifts by $\Delta_j \to \Delta+1$ in agreement with the operator products between $\O_{\Delta_j}$ and the respective currents~\cite{Donnay:2018neh,Himwich:2019dug}.

\section{Modeling the Conformally Soft Sector}
 \label{sec:conf_soft_sec}
 
 In this section we combine the ingredients for the conformally soft sector. We speculate on the implications for non-trivial levels and central charges in celestial CFT in section~\ref{sec:cent_ex} and then discuss 2D effective actions for the conformally soft modes in section~\ref{sec:2D_EFT}.

\subsection{Central Extensions} 
\label{sec:cent_ex}

The operator products between the 2D operators $\{\J,\S,\N,\C,\T^A,\Y^A\}$ at the top of the celestial Goldstone and memory diamonds give rise to levels and central extensions when they are non-vanishing.
For the leading soft photon and graviton we have the following two-point functions~\cite{Nande:2017dba,Himwich:2020rro,Arkani-Hamed:2020gyp}\footnote{The definition of 2D operators via the inner product~\eqref{eq:2Dop} between the bulk fields and generalized conformal primary wavefunctions should guarantee the necessary SL(2,$\mathbb{C}$) transformation properties. While this may not be obvious from e.g.~\eqref{corrJS} we anticipate that we can use the operator $\widehat{\omega}$ defined in~\eqref{omegahat} to write a two-point function
\be
 \langle i\S(w_i,\bw_i)i\S(w_j,\bw_j)\rangle \mapsto k_{\S\S}\log\Big(|w_{ij}|^2 \widehat{\omega}_i \widehat{\omega}_j\Big)
\ee
which is SL(2,$\mathbb{C}$) invariant since $\omega\mapsto\omega|cw+d|^2$ and $w-w'\mapsto\frac{w-w'}{(cw+d)(cw'+d)}$.  Note that the energy operators are replacing the cutoff $\mu$ which appeared in~\cite{Haehl:2019eae}, and was interpreted there as signaling a conformal anomaly.}
 \begin{equation}\label{corrJS}
\badat{3}
    & \langle \J(w_i,\bw_i)\J(w_j,\bw_j)\rangle= k_{\J\J} \log|w_{ij}|^2\,,\quad   \langle \N(w_i,\bw_i)\N(w_j,\bw_j)\rangle={\frac{1}{4}}k_{\N\N} |w_{ij}|^2\log|w_{ij}|^2\,,\\
    & \langle \J(w_i,\bw_i){i}\S(w_j,\bw_j)\rangle= k_{\J\S}\log|w_{ij}|^2\,,\quad  \langle \N(w_i,\bw_i){i}\C(w_j,\bw_j)\rangle={\frac{1}{4}} k_{\N\C}|w_{ij}|^2\log|w_{ij}|^2\,,\\
    & \langle {i}\S(w_i,\bw_i){i}\S(w_j,\bw_j)\rangle= k_{\S\S} \log|w_{ij}|^2\,,\quad    \langle {i}\C(w_i,\bw_i){i}\C(w_j,\bw_j)\rangle={\frac{1}{4}}k_{\C\C}\, |w_{ij}|^2\log|w_{ij}|^2\,,
\eadat
\end{equation}
while for the sub-leading soft graviton we have~\cite{Ball:2019atb,Haehl:2019eae}
 \begin{equation}\label{corrTSHT}
 \badat{3}
    \langle \T^{w_i}(w_i,\bw_i)\T^{w_j}(w_j,\bw_j)\rangle&={\frac{3}{2}} c_{\T\T} w_{ij}^2\log|w_{ij}|^2\,, \\
      \langle \T^{w_i}(w_i,\bw_i)\Y^{w_j}(w_j,\bw_j)\rangle&={\frac{3}{2}}  c_{\T\Y}w_{ij}^2\log|w_{ij}|^2\,, \\
        \langle \Y^{w_i}(w_i,\bw_i)\Y^{w_j}(w_j,\bw_j)\rangle&={\frac{3}{2}}  c_{\Y\Y} w_{ij}^2\log|w_{ij}|^2\,, 
\eadat
\end{equation}
and analogous expressions for the opposite helicity modes.
In the above expressions we have allowed for various constants that are fixed by the dynamics and correspond to levels and central charges. From multi-soft limits~\cite{He:2015zea,Fotopoulos:2019vac} there is evidence that\footnote{See~\cite{Nande:2017dba} for an example where this can get modified when the currents are complexified in the presence of magnetic monopole sources. Consequences of generalizing this have also been examined in~\cite{Fan:2020xjj}.
}
\be
k_{\J\J}=0\,, \quad k_{\N\N}=0\,, \quad c_{\T\T}=0\,,
\ee
at least at tree-level while canonical normalization implies~\cite{Nande:2017dba,Arkani-Hamed:2020gyp,Ball:2019atb}
\begin{equation}
    k_{\J\S}=1\,, \quad k_{\N\C}=1\,, \quad c_{\T\Y}=1\,.
\end{equation}
Non-trivial levels in the two-point functions for the Goldstone bosons arise from loop effects. These are cleanly captured in the eikonal approximation where the celestial amplitude factorizes into a soft and a hard part
\be
 {\cal A}={\cal A}_{soft}{\cal A}_{hard}\,,
\ee
at least at leading order in the soft expansion. The levels for the Goldstone bosons corresponding to the leading conformally soft photon and graviton diamonds are given by the cusp anomalous dimensions~\cite{Arkani-Hamed:2020gyp}\footnote{See~\cite{Gonzalez:2020tpi,Magnea:2021fvy,Gonzalez:2021dxw} for recent results on the IR structure of celestial loop amplitudes in non-Abelian gauge theory.}
\be
k_{\S\S}={-}\frac{e^2}{4\pi^2}\log\Lambda_{IR}\,, \quad k_{\C\C}={\frac{\kappa^2}{2\pi^2}}\log\Lambda_{IR}\,.
\ee
Dressing the bare hard operators with the Goldstone boson $\S$ and $\C$ in~\eqref{Wefinal} and~\eqref{Wgfinal} leads to a cancellation of the IR divergent soft factor ${\cal A}_{soft}$ such that the dressed celestial amplitude is IR finite and given by ${\cal A}_{hard}$.~\cite{Arkani-Hamed:2020gyp}

On the other hand, the sub-leading soft dressings of~\cite{Choi:2019rlz}, recast as conformally soft dressings in~\eqref{Wefinal} and~\eqref{Wgfinal}, 
are only valid up to leading order in the couplings. In particular, loop corrections mix the leading and sub-leading conformally soft contributions. Nevertheless, one could try to reverse engineer the sub-leading dressings so as to remove certain collinear divergences between hard operators and the respective current. For a bare operator $\mathcal{O}_{h_j,\bar h_j}$, its OPE with the stress tensor takes the form~\cite{Fotopoulos:2019tpe}
\be
\T_{ww}{\cal O}_{h_j,\bar h_j}\sim~\frac{h_j {\cal O}_{h_j,\bar h_j}}{(w-w_j)^2}+ \frac{\p_{w_j} {\cal O}_{h_j,\bar h_j}}{w-w_j} \,,
\ee
where we have dropped regular terms. The sub-leading conformally soft dressing we found in~\eqref{Wgfinal} is expected to imply the OPE
\be
\T_{ww}:e^{  2[h_j \p_{w_j}\Y^{w_j}+\Y^{w_j}\p_{w_j}]}{\cal O}_{h_j,\bar h_j}:\sim \mathrm{regular}\,.
\ee
The one-loop exact correction to the energy momentum tensor~\cite{He:2017fsb} obtained in dimensional regularization ($D=4-\epsilon$) induces the following shift in the parent primaries
\be
\frac{1}{3!}\p_z^3\Delta\T^z=\frac{1}{\pi\kappa^2\epsilon}\int d^2w \frac{1}{z-w}(3\p_w^2\N\p_w\p_\bw^2 \N+\p_w^3 \N\p_\bw^2 \N)\,.
\ee
This shift in the charges might be expected given that superrotations and supertranslations do not commute and that the soft charges should reproduce this algebra~\cite{Barnich:2011ct}.

\subsection{Effective 2D Descriptions}\label{sec:2D_EFT}
We close this paper with a discussion of the role of the operators at the top of the celestial diamonds. These modes govern the spontaneous symmetry breaking of asymptotic symmetries and should be described by some intrinsically 2D effective theory.  Various elements of this theory have already been explored in the celestial CFT literature.  We summarize them here, then review the simplest example in enough detail to extract what the general features are for the free limit, before giving a simple toy model of 2D effective actions for the modes at the top of our celestial diamonds. Its completion into a non-linear model is an interesting problem which we leave to future work.

\paragraph{Leading Soft Photon/Gluon} The spontaneous breaking of large gauge symmetries in QED is captured by a free boson in 2D~\cite{Nande:2017dba}.  Vertex operators constructed from this boson give the dressing for charged states. In Yang-Mills this gets promoted to a Lie algebra valued free boson~\cite{Magnea:2021fvy,Gonzalez:2021dxw}.  The interacting version~\cite{Cheung:2016iub} is described by a Chern-Simons theory on a hyperbolic slice of Minkowski space~\cite{deBoer:2003vf}, which should be dual to a Wess-Zumino-Witten model. 

\paragraph{Leading Soft Graviton} While an understanding of the soft phase space and eikonal factorizations have much earlier origins, the main precedent for how we are viewing the leading gravitational dressings in celestial CFT are the references~\cite{Himwich:2020rro,Arkani-Hamed:2020gyp}. There the IR divergences are captured by correlators of vertex operators built from the supertranslation Goldstone mode. A higher derivative 2D effective action is given in~\cite{Kalyanapuram:2020epb,Kalyanapuram:2021bvf}.

\paragraph{Sub-leading Soft Graviton} An effective action for superrotations was constructed in~\cite{Nguyen:2020hot}.  As within a wider 2D CFT context~\cite{Haehl:2019eae,Nguyen:2021dpa}, the spontaneous breaking of Diff$(S^2)$ to Virasoro is governed by an Alekseev-Shatashvili action~\cite{Alekseev:1988ce}.

\vspace{1em}

The leading conformally soft photon is the simplest of the above examples. The claim~\cite{Cheung:2016iub,Nande:2017dba} is that the spontaneous symmetry breaking of large U(1) gauge symmetries is governed by a free scalar in a 2D CFT. Let us demonstrate how this 2D toy model has the same structure as the celestial diamond for the leading conformally soft theorem.

For the theory of a free boson $S=\int d^2w \p_w \phi \p_{\bw}  \phi$.  Taking the field $\phi(w,\bw)$ to lie at the top of the diamond, we see that it
descends to primary operators at the left and right corners.  These, in turn, descend to an operator at the bottom which encodes a `shortening' condition of the multiplet.  In the classification of~\cite{Pasterski:2021fjn} the fields $\p_w \phi$ and $\p_{\wb} \phi$ are type Ia and Ib primary descendants.  They play the role of currents associated to the symmetry $\phi \to \phi + const$ of the action.  By taking an extra derivative we obtain the type II primary descendant $\p_w \p_{\wb} \phi$ which is zero by the equations of motion (considered as an operator equation valid in correlation functions up to contact terms)
and defines the conservation equation for the currents which can, in turn, be integrated to define charges. While $\phi$ does not transform as a primary,\footnote{Note that $\phi$ itself is not a conformal primary in the sense that e.g. its two point function takes a logarithmic form which does not scale covariantly. Indeed, under the state-operator map,  $\phi$ cannot be associated to a normalizable state of the Hilbert space. On the other hand the OPE of $\phi$ with the stress tensor is the one characteristic of a primary operator with $h=\bar{h}=0$.  Therefore the generators $L_i$ and $\bar L_i$ act on it as on usual primary fields, e.g. $[L_1,\phi(0)]=0$. This means that the purely algebraic considerations of section~3 of~\cite{Pasterski:2021fjn} hold also for $\phi$ (as if it were a primary). } it is a building block for constructing vertex operators  $V_\alpha(w,\bw)=:e^{i \alpha \phi(w,\bw)}:$ which are well defined primaries with conformal dimension $\alpha^2$, charged under the symmetry generated by  $\partial_w \phi$ and $ \partial_\bw \phi$. These can be used to `dress' the vacuum as $V_\alpha(0)|0\rangle \equiv |\alpha \rangle$, which now carries a charge.  Conversely, given a charged state of the Fock space, we can make it neutral by acting on the vacuum with an opportune $V_\alpha$.  

We thus see that this toy model has circled back to the spontaneous symmetry breaking story that initiated our trek into the conformally soft sector.  For this example, $\phi$ corresponds to $\cal{O}^1_{0,0}$ at the top of the leading conformally soft photon diamond. Let us now try to construct analogous toy models for the free limit of ${\cal O}^s_{\Delta,J}$.  The main takeaways are the shortening conditions, the currents, and the vertex operators.

An interesting toy model that captures these features of the diamond structure can be formally obtained by considering a higher derivative Gaussian theory with action\footnote{In the previous sections we have been discussing the Goldstone and memory modes separately. Here we are sidestepping how they interplay in the 2D picture.  In the bulk we know that they are symplectic partners and inherit Hermiticity conditions from the reality of the gauge field and metric. This might hint at a single dual field, however our understanding of the 2D Hilbert space is still evolving~\cite{Fan:2021isc,Crawley:2021ivb}.}  
\begin{equation}
\label{Stoy}
  S=\int d^2w \left[\partial_w^{\n}\mathcal{O}^s_{\Delta, J}\partial_{\bar w}^{\bar \n}\mathcal{O}^s_{\Delta, J} +\partial_{\bar w}^{\n}\mathcal{O}^s_{\Delta, -J}\partial_{ w}^{\bar \n}\mathcal{O}^s_{\Delta, -J}\right] \, ,
\end{equation}
where, in order to have a scale invariant action, we choose operators $\mathcal{O}$ with dimension 
$\Delta=1-\frac{\n + \bar \n}{2}$ and spin $J=\frac{\bar \n - \n}{2}$ where $k,\bar k \in \mathbb{Z}_{>}$.
Here we have introduced the label $s$ in order to make contact with the 4D notation of~\eqref{eq:2Dop}. It is fixed to $\n + \bar \n=2s$, consistent with $\Delta=1-s$ for operators at the top of the celestial diamonds~\cite{Pasterski:2021fjn}. The  equations of motion can be written as follows
\begin{equation}
 \partial_w^{\n} \partial_{\bar w}^{\bar \n}  \mathcal{O}^s_{\Delta, J} =0\, .
\end{equation}
We notice this takes us to the bottom corner of the diamond. By asking that the two point function of $\mathcal{O}_{\Delta J}$ is the corresponding Green's function we obtain
\begin{equation}
\label{2ptOtoy}
    \langle\mathcal{O}^s_{\Delta, J}(w,\bar w)\mathcal{O}^s_{\Delta, J}(0,0) \rangle 
    \propto
    w^{\n-1} \bar w^{\bar \n-1} \log(w \bar w) \, .
\end{equation}
This implies that we can define the two operators 
\begin{equation}
  \mathcal{J}_L\equiv \partial_{\bar w}^{\bar \n}  \mathcal{O}^s_{\Delta, J} \, ,
  \qquad
    \mathcal{J}_R \equiv \partial_w^{\n} \mathcal{O}^s_{\Delta, J} \, ,
\end{equation}
which lie at the left and right corners of the diamond and thus have 2D spin $\pm s$. 
These currents have canonical two point functions (without logs) 
 \begin{equation}
 \langle\mathcal{J}_L(w,\bar w) \mathcal{J}_L(0,0) \rangle  \propto
 ( w)^{\n-1} (\bar w)^{-\bar \n-1}\, ,
     \qquad 
     \langle\mathcal{J}_R(w,\bar w) \mathcal{J}_R(0,0) \rangle  \propto 
     ( w)^{- \n-1} (\bar w)^{\bar \n-1}\, ,
     \end{equation}
  while $\langle\mathcal{J}_L(w,\bar w) \mathcal{J}_R(0,0) \rangle$ reduces to a contact term.
The $k-1$th derivatives in $w$ of $\J_L$ gives antiholomorphic generalized currents in the sense of section~\ref{sec:gen_j} (and similarly for $\J_R$).\footnote{ The $\J_L$ are polynomial in $w$. In the recent investigations~\cite{Banerjee:2020zlg,Banerjee:2020vnt} the coefficients of these polynomials are shown to obey interesting symmetry algebras.} 

We thus see that this toy model gives us shortening conditions and currents, which replicate the structure of soft charges and generalized celestial currents in section~\ref{sec:softcharges}. Moreover, we expect the associated vertex operators to match the conformally soft dressings we constructed in section~\ref{sec:dressing}. It would be very interesting to systematically study these theories and find how to define, from first principles, their vertex operators. In practice, this task should be straightforward since  we are dealing with a special class of generalized free theories where all correlation functions can be obtained by Wick contractions. It is worth pointing out though that, while being simple, these theories are of a subtle type. Besides the fact that the Gaussian fields have logarithmic two point functions~\eqref{2ptOtoy}, in general these theories do not have a stress tensor (apart from simple cases e.g. $\n=\bar \n=1$ which corresponds to the free boson theory), therefore they are not  full fledged local CFTs. 

Our toy model captures the free limit of the examples from celestial CFT discussed above. Generalizations of this model with operators $\O$ that transform under a global symmetry can be easily considered. Moreover, knowing how the Yang Mills~\cite{Cheung:2016iub} and sub-leading gravity~\cite{Nguyen:2020hot} get completed to interesting non-linear models, invites us to look for analogs for the other conformally soft modes.  While the most sub-leading soft theorems do not have an obvious spontaneous symmetry breaking interpretation, the associated currents are powerful enough to fix the OPE~\cite{Pate:2019lpp}. Our investigations into celestial diamonds have handed us the ingredients for the conformally soft sector, which we expect to be a rich microcosm of celestial physics.

\section*{Acknowledgements}

We would like to thank Sangmin Choi, Raoul Santachiara, Shu-Heng Shao, Herman Verlinde, and Bernardo Zan for useful discussions. The work of S.P. is supported by the Princeton Center for Theoretical Science. The work of A.P. and E.T. is supported by the European Research Council (ERC) under the European Union’s Horizon 2020 research and innovation programme (grant agreement No 852386).

\bibliographystyle{utphys}
\bibliography{NullStates}

\providecommand{\href}[2]{#2}\begingroup\raggedright\begin{thebibliography}{10}

\bibitem{Pasterski:2017ylz}
S.~Pasterski, S.-H. Shao, and A.~Strominger, ``{Gluon Amplitudes as 2d
  Conformal Correlators},''
  \href{http://dx.doi.org/10.1103/PhysRevD.96.085006}{{\em Phys. Rev.}
  {\bfseries D96} no.~8, (2017) 085006},
\href{http://arxiv.org/abs/1706.03917}{{\ttfamily arXiv:1706.03917 [hep-th]}}.

\bibitem{Donnay:2018neh}
L.~Donnay, A.~Puhm, and A.~Strominger, ``{Conformally Soft Photons and
  Gravitons},'' \href{http://dx.doi.org/10.1007/JHEP01(2019)184}{{\em JHEP}
  {\bfseries 01} (2019) 184},
\href{http://arxiv.org/abs/1810.05219}{{\ttfamily arXiv:1810.05219 [hep-th]}}.

\bibitem{Pasterski:2020pdk}
S.~Pasterski and A.~Puhm, ``{Shifting Spin on the Celestial Sphere},''
  \href{http://arxiv.org/abs/2012.15694}{{\ttfamily arXiv:2012.15694
  [hep-th]}}.

\bibitem{deBoer:2003vf}
J.~de~Boer and S.~N. Solodukhin, ``{A Holographic reduction of Minkowski
  space-time},'' \href{http://dx.doi.org/10.1016/S0550-3213(03)00494-2}{{\em
  Nucl. Phys.} {\bfseries B665} (2003) 545--593},
\href{http://arxiv.org/abs/hep-th/0303006}{{\ttfamily arXiv:hep-th/0303006
  [hep-th]}}.

\bibitem{Cheung:2016iub}
C.~Cheung, A.~de~la Fuente, and R.~Sundrum, ``{4D scattering amplitudes and
  asymptotic symmetries from 2D CFT},''
  \href{http://dx.doi.org/10.1007/JHEP01(2017)112}{{\em JHEP} {\bfseries 01}
  (2017) 112},
\href{http://arxiv.org/abs/1609.00732}{{\ttfamily arXiv:1609.00732 [hep-th]}}.

\bibitem{Haehl:2019eae}
F.~M. Haehl, W.~Reeves, and M.~Rozali, ``{Reparametrization modes, shadow
  operators, and quantum chaos in higher-dimensional CFTs},''
  \href{http://dx.doi.org/10.1007/JHEP11(2019)102}{{\em JHEP} {\bfseries 11}
  (2019) 102}, \href{http://arxiv.org/abs/1909.05847}{{\ttfamily
  arXiv:1909.05847 [hep-th]}}.

\bibitem{Nguyen:2020hot}
K.~Nguyen and J.~Salzer, ``{The effective action of superrotation modes},''
  \href{http://dx.doi.org/10.1007/JHEP02(2021)108}{{\em JHEP} {\bfseries 02}
  (2021) 108}, \href{http://arxiv.org/abs/2008.03321}{{\ttfamily
  arXiv:2008.03321 [hep-th]}}.

\bibitem{Nguyen:2021dpa}
K.~Nguyen, ``{Reparametrization modes in 2d CFT and the effective theory of
  stress tensor exchanges},'' \href{http://arxiv.org/abs/2101.08800}{{\ttfamily
  arXiv:2101.08800 [hep-th]}}.

\bibitem{Campiglia:2014yka}
M.~Campiglia and A.~Laddha, ``{Asymptotic symmetries and subleading soft
  graviton theorem},'' \href{http://dx.doi.org/10.1103/PhysRevD.90.124028}{{\em
  Phys. Rev.} {\bfseries D90} no.~12, (2014) 124028},
\href{http://arxiv.org/abs/1408.2228}{{\ttfamily arXiv:1408.2228 [hep-th]}}.

\bibitem{Compere:2018ylh}
G.~Comp\`ere, A.~Fiorucci, and R.~Ruzziconi, ``{Superboost transitions,
  refraction memory and super-Lorentz charge algebra},''
  \href{http://dx.doi.org/10.1007/JHEP11(2018)200}{{\em JHEP} {\bfseries 11}
  (2018) 200}, \href{http://arxiv.org/abs/1810.00377}{{\ttfamily
  arXiv:1810.00377 [hep-th]}}.

\bibitem{Donnay:2020guq}
L.~Donnay, S.~Pasterski, and A.~Puhm, ``{Asymptotic Symmetries and Celestial
  CFT},'' \href{http://dx.doi.org/10.1007/JHEP09(2020)176}{{\em JHEP}
  {\bfseries 09} (2020) 176}, \href{http://arxiv.org/abs/2005.08990}{{\ttfamily
  arXiv:2005.08990 [hep-th]}}.

\bibitem{Pasterski:2021fjn}
S.~Pasterski, A.~Puhm, and E.~Trevisani, ``{Celestial Diamonds: Conformal
  Multiplets in Celestial CFT},''
  \href{http://arxiv.org/abs/2105.03516}{{\ttfamily arXiv:2105.03516
  [hep-th]}}.

\bibitem{Banerjee:2018gce}
S.~Banerjee, ``{Null Infinity and Unitary Representation of The Poincare
  Group},'' \href{http://dx.doi.org/10.1007/JHEP01(2019)205}{{\em JHEP}
  {\bfseries 01} (2019) 205}, \href{http://arxiv.org/abs/1801.10171}{{\ttfamily
  arXiv:1801.10171 [hep-th]}}.

\bibitem{Banerjee:2018fgd}
S.~Banerjee, ``{Symmetries of free massless particles and soft theorems},''
  \href{http://dx.doi.org/10.1007/s10714-019-2609-z}{{\em Gen. Rel. Grav.}
  {\bfseries 51} no.~9, (2019) 128},
  \href{http://arxiv.org/abs/1804.06646}{{\ttfamily arXiv:1804.06646
  [hep-th]}}.

\bibitem{Banerjee:2019aoy}
S.~Banerjee, P.~Pandey, and P.~Paul, ``{Conformal properties of soft operators:
  Use of null states},''
  \href{http://dx.doi.org/10.1103/PhysRevD.101.106014}{{\em Phys. Rev. D}
  {\bfseries 101} no.~10, (2020) 106014},
  \href{http://arxiv.org/abs/1902.02309}{{\ttfamily arXiv:1902.02309
  [hep-th]}}.

\bibitem{Banerjee:2019tam}
S.~Banerjee and P.~Pandey, ``{Conformal properties of soft-operators. Part II.
  Use of null-states},'' \href{http://dx.doi.org/10.1007/JHEP02(2020)067}{{\em
  JHEP} {\bfseries 02} (2020) 067},
  \href{http://arxiv.org/abs/1906.01650}{{\ttfamily arXiv:1906.01650
  [hep-th]}}.

\bibitem{Arkani-Hamed:2020gyp}
N.~Arkani-Hamed, M.~Pate, A.-M. Raclariu, and A.~Strominger, ``{Celestial
  Amplitudes from UV to IR},''
  \href{http://arxiv.org/abs/2012.04208}{{\ttfamily arXiv:2012.04208
  [hep-th]}}.

\bibitem{Strominger:2013jfa}
A.~Strominger, ``{On BMS Invariance of Gravitational Scattering},''
  \href{http://dx.doi.org/10.1007/JHEP07(2014)152}{{\em JHEP} {\bfseries 07}
  (2014) 152},
\href{http://arxiv.org/abs/1312.2229}{{\ttfamily arXiv:1312.2229 [hep-th]}}.

\bibitem{Choi:2019rlz}
S.~Choi and R.~Akhoury, ``{Subleading soft dressings of asymptotic states in
  QED and perturbative quantum gravity},''
  \href{http://dx.doi.org/10.1007/JHEP09(2019)031}{{\em JHEP} {\bfseries 09}
  (2019) 031}, \href{http://arxiv.org/abs/1907.05438}{{\ttfamily
  arXiv:1907.05438 [hep-th]}}.

\bibitem{Ball:2019atb}
A.~Ball, E.~Himwich, S.~A. Narayanan, S.~Pasterski, and A.~Strominger,
  ``{Uplifting AdS$_{3}$/CFT$_{2}$ to flat space holography},''
  \href{http://dx.doi.org/10.1007/JHEP08(2019)168}{{\em JHEP} {\bfseries 08}
  (2019) 168}, \href{http://arxiv.org/abs/1905.09809}{{\ttfamily
  arXiv:1905.09809 [hep-th]}}.

\bibitem{Pasterski:2017kqt}
S.~Pasterski and S.-H. Shao, ``{Conformal basis for flat space amplitudes},''
  \href{http://dx.doi.org/10.1103/PhysRevD.96.065022}{{\em Phys. Rev.}
  {\bfseries D96} no.~6, (2017) 065022},
\href{http://arxiv.org/abs/1705.01027}{{\ttfamily arXiv:1705.01027 [hep-th]}}.

\bibitem{Ashtekar:1987tt}
A.~Ashtekar, {\em {Asymptotic Quantization: Based on 1984 Naples Lectures}}.
\newblock Naples, Italy: Bibliopolis 107 P. (Monographs and Textbooks in
  Physical Science, 2),
1987.
\newblock

\bibitem{Crnkovic:1986ex}
C.~{Crnkovic} and E.~{Witten}, ``{Covariant description of canonical formalism
  in geometrical theories},'' in {\em Three Hundred Years of Gravitation},
  pp.~676--684.
\newblock {S.~W. Hawking} and { W. Israel}, 1987.

\bibitem{Lee:1990nz}
J.~Lee and R.~M. Wald, ``{Local symmetries and constraints},''
\href{http://dx.doi.org/10.1063/1.528801}{{\em J. Math. Phys.} {\bfseries 31}
  (1990) 725--743}.

\bibitem{Wald:1999wa}
R.~M. Wald and A.~Zoupas, ``{A General definition of `conserved quantities' in
  general relativity and other theories of gravity},''
  \href{http://dx.doi.org/10.1103/PhysRevD.61.084027}{{\em Phys. Rev.}
  {\bfseries D61} (2000) 084027},
\href{http://arxiv.org/abs/gr-qc/9911095}{{\ttfamily arXiv:gr-qc/9911095
  [gr-qc]}}.

\bibitem{Weinberg:1995mt}
S.~Weinberg, {\em {The Quantum theory of fields. Vol. 1: Foundations}}.
\newblock Cambridge University Press, 6, 2005.

\bibitem{Strominger:2017zoo}
A.~Strominger, {\em {Lectures on the Infrared Structure of Gravity and Gauge
  Theory}}.
\newblock {Princeton University Press}, 2018.
\newblock
\href{http://arxiv.org/abs/1703.05448}{{\ttfamily arXiv:1703.05448 [hep-th]}}.
\newblock

\bibitem{Barnich:2021dta}
G.~Barnich and R.~Ruzziconi, ``{Coadjoint representation of the BMS group on
  celestial Riemann surfaces},''
  \href{http://arxiv.org/abs/2103.11253}{{\ttfamily arXiv:2103.11253 [gr-qc]}}.

\bibitem{He:2014cra}
T.~He, P.~Mitra, A.~P. Porfyriadis, and A.~Strominger, ``{New Symmetries of
  Massless QED},'' \href{http://dx.doi.org/10.1007/JHEP10(2014)112}{{\em JHEP}
  {\bfseries 10} (2014) 112},
\href{http://arxiv.org/abs/1407.3789}{{\ttfamily arXiv:1407.3789 [hep-th]}}.

\bibitem{He:2014laa}
T.~He, V.~Lysov, P.~Mitra, and A.~Strominger, ``{BMS supertranslations and
  Weinberg's soft graviton theorem},''
  \href{http://dx.doi.org/10.1007/JHEP05(2015)151}{{\em JHEP} {\bfseries 05}
  (2015) 151},
\href{http://arxiv.org/abs/1401.7026}{{\ttfamily arXiv:1401.7026 [hep-th]}}.

\bibitem{Lysov:2014csa}
V.~Lysov, S.~Pasterski, and A.~Strominger, ``{Low's Subleading Soft Theorem as
  a Symmetry of QED},''
  \href{http://dx.doi.org/10.1103/PhysRevLett.113.111601}{{\em Phys. Rev.
  Lett.} {\bfseries 113} no.~11, (2014) 111601},
\href{http://arxiv.org/abs/1407.3814}{{\ttfamily arXiv:1407.3814 [hep-th]}}.

\bibitem{Kapec:2014opa}
D.~Kapec, V.~Lysov, S.~Pasterski, and A.~Strominger, ``{Semiclassical Virasoro
  symmetry of the quantum gravity $ \mathcal{S}$-matrix},''
  \href{http://dx.doi.org/10.1007/JHEP08(2014)058}{{\em JHEP} {\bfseries 08}
  (2014) 058},
\href{http://arxiv.org/abs/1406.3312}{{\ttfamily arXiv:1406.3312 [hep-th]}}.

\bibitem{He:2015zea}
T.~He, P.~Mitra, and A.~Strominger, ``{2D Kac-Moody Symmetry of 4D Yang-Mills
  Theory},'' \href{http://dx.doi.org/10.1007/JHEP10(2016)137}{{\em JHEP}
  {\bfseries 10} (2016) 137},
\href{http://arxiv.org/abs/1503.02663}{{\ttfamily arXiv:1503.02663 [hep-th]}}.

\bibitem{Himwich:2019dug}
E.~Himwich and A.~Strominger, ``{Celestial Current Algebra from Low's
  Subleading Soft Theorem},''
\href{http://arxiv.org/abs/1901.01622}{{\ttfamily arXiv:1901.01622 [hep-th]}}.

\bibitem{Kapec:2016jld}
D.~Kapec, P.~Mitra, A.-M. Raclariu, and A.~Strominger, ``{2D Stress Tensor for
  4D Gravity},'' \href{http://dx.doi.org/10.1103/PhysRevLett.119.121601}{{\em
  Phys. Rev. Lett.} {\bfseries 119} no.~12, (2017) 121601},
\href{http://arxiv.org/abs/1609.00282}{{\ttfamily arXiv:1609.00282 [hep-th]}}.

\bibitem{Strominger:2014pwa}
A.~Strominger and A.~Zhiboedov, ``{Gravitational Memory, BMS Supertranslations
  and Soft Theorems},'' \href{http://dx.doi.org/10.1007/JHEP01(2016)086}{{\em
  JHEP} {\bfseries 01} (2016) 086},
  \href{http://arxiv.org/abs/1411.5745}{{\ttfamily arXiv:1411.5745 [hep-th]}}.

\bibitem{Pasterski:2015tva}
S.~Pasterski, A.~Strominger, and A.~Zhiboedov, ``{New Gravitational
  Memories},'' \href{http://dx.doi.org/10.1007/JHEP12(2016)053}{{\em JHEP}
  {\bfseries 12} (2016) 053}, \href{http://arxiv.org/abs/1502.06120}{{\ttfamily
  arXiv:1502.06120 [hep-th]}}.

\bibitem{Dolan:2001ih}
L.~Dolan, C.~R. Nappi, and E.~Witten, ``{Conformal operators for partially
  massless states},''
  \href{http://dx.doi.org/10.1088/1126-6708/2001/10/016}{{\em JHEP} {\bfseries
  10} (2001) 016}, \href{http://arxiv.org/abs/hep-th/0109096}{{\ttfamily
  arXiv:hep-th/0109096}}.

\bibitem{Brust:2016gjy}
C.~Brust and K.~Hinterbichler, ``{Free \ensuremath{\square}$^{k}$ scalar
  conformal field theory},''
  \href{http://dx.doi.org/10.1007/JHEP02(2017)066}{{\em JHEP} {\bfseries 02}
  (2017) 066}, \href{http://arxiv.org/abs/1607.07439}{{\ttfamily
  arXiv:1607.07439 [hep-th]}}.

\bibitem{Fotopoulos:2019vac}
A.~Fotopoulos, S.~Stieberger, T.~R. Taylor, and B.~Zhu, ``{Extended BMS Algebra
  of Celestial CFT},'' \href{http://dx.doi.org/10.1007/JHEP03(2020)130}{{\em
  JHEP} {\bfseries 03} (2020) 130},
  \href{http://arxiv.org/abs/1912.10973}{{\ttfamily arXiv:1912.10973
  [hep-th]}}.

\bibitem{Himwich:2019qmj}
E.~Himwich, Z.~Mirzaiyan, and S.~Pasterski, ``{A Note on the Subleading Soft
  Graviton},''
\href{http://arxiv.org/abs/1902.01840}{{\ttfamily arXiv:1902.01840 [hep-th]}}.

\bibitem{Stieberger:2018onx}
S.~Stieberger and T.~R. Taylor, ``{Symmetries of Celestial Amplitudes},''
  \href{http://dx.doi.org/10.1016/j.physletb.2019.03.063}{{\em Phys. Lett. B}
  {\bfseries 793} (2019) 141--143},
  \href{http://arxiv.org/abs/1812.01080}{{\ttfamily arXiv:1812.01080
  [hep-th]}}.

\bibitem{Nande:2017dba}
A.~Nande, M.~Pate, and A.~Strominger, ``{Soft Factorization in QED from 2D
  Kac-Moody Symmetry},'' \href{http://dx.doi.org/10.1007/JHEP02(2018)079}{{\em
  JHEP} {\bfseries 02} (2018) 079},
\href{http://arxiv.org/abs/1705.00608}{{\ttfamily arXiv:1705.00608 [hep-th]}}.

\bibitem{Himwich:2020rro}
E.~Himwich, S.~A. Narayanan, M.~Pate, N.~Paul, and A.~Strominger, ``{The Soft
  $\mathcal{S}$-Matrix in Gravity},''
  \href{http://dx.doi.org/10.1007/JHEP09(2020)129}{{\em JHEP} {\bfseries 09}
  (2020) 129}, \href{http://arxiv.org/abs/2005.13433}{{\ttfamily
  arXiv:2005.13433 [hep-th]}}.

\bibitem{Fan:2020xjj}
W.~Fan, A.~Fotopoulos, S.~Stieberger, and T.~R. Taylor, ``{On Sugawara
  construction on Celestial Sphere},''
  \href{http://dx.doi.org/10.1007/JHEP09(2020)139}{{\em JHEP} {\bfseries 09}
  (2020) 139}, \href{http://arxiv.org/abs/2005.10666}{{\ttfamily
  arXiv:2005.10666 [hep-th]}}.

\bibitem{Gonzalez:2020tpi}
H.~A. Gonz\'alez, A.~Puhm, and F.~Rojas, ``{Loop corrections to celestial
  amplitudes},'' \href{http://dx.doi.org/10.1103/PhysRevD.102.126027}{{\em
  Phys. Rev. D} {\bfseries 102} no.~12, (2020) 126027},
  \href{http://arxiv.org/abs/2009.07290}{{\ttfamily arXiv:2009.07290
  [hep-th]}}.

\bibitem{Magnea:2021fvy}
L.~Magnea, ``{Non-abelian infrared divergences on the celestial sphere},''
  \href{http://arxiv.org/abs/2104.10254}{{\ttfamily arXiv:2104.10254
  [hep-th]}}.

\bibitem{Gonzalez:2021dxw}
H.~A. Gonz\'alez and F.~Rojas, ``{The structure of IR divergences in celestial
  gluon amplitudes},'' \href{http://arxiv.org/abs/2104.12979}{{\ttfamily
  arXiv:2104.12979 [hep-th]}}.

\bibitem{Fotopoulos:2019tpe}
A.~Fotopoulos and T.~R. Taylor, ``{Primary Fields in Celestial CFT},''
  \href{http://dx.doi.org/10.1007/JHEP10(2019)167}{{\em JHEP} {\bfseries 10}
  (2019) 167},
\href{http://arxiv.org/abs/1906.10149}{{\ttfamily arXiv:1906.10149 [hep-th]}}.

\bibitem{He:2017fsb}
T.~He, D.~Kapec, A.-M. Raclariu, and A.~Strominger, ``{Loop-Corrected Virasoro
  Symmetry of 4D Quantum Gravity},''
  \href{http://dx.doi.org/10.1007/JHEP08(2017)050}{{\em JHEP} {\bfseries 08}
  (2017) 050},
\href{http://arxiv.org/abs/1701.00496}{{\ttfamily arXiv:1701.00496 [hep-th]}}.

\bibitem{Barnich:2011ct}
G.~Barnich and C.~Troessaert, ``{Supertranslations call for superrotations},''
  {\em PoS} {\bfseries CNCFG} (2010) 010,
  \href{http://arxiv.org/abs/1102.4632}{{\ttfamily arXiv:1102.4632 [gr-qc]}}.
[Ann. U. Craiova Phys.21,S11(2011)].

\bibitem{Kalyanapuram:2020epb}
N.~Kalyanapuram, ``{Soft Gravity by Squaring Soft QED on the Celestial
  Sphere},'' \href{http://dx.doi.org/10.1103/PhysRevD.103.085016}{{\em Phys.
  Rev. D} {\bfseries 103} no.~8, (2021) 085016},
  \href{http://arxiv.org/abs/2011.11412}{{\ttfamily arXiv:2011.11412
  [hep-th]}}.

\bibitem{Kalyanapuram:2021bvf}
N.~Kalyanapuram, ``{The Holographic Soft $S$-Matrix in QED and Gravity},''
  \href{http://arxiv.org/abs/2105.04314}{{\ttfamily arXiv:2105.04314
  [hep-th]}}.

\bibitem{Alekseev:1988ce}
A.~Alekseev and S.~L. Shatashvili, ``{Path Integral Quantization of the
  Coadjoint Orbits of the Virasoro Group and 2D Gravity},''
  \href{http://dx.doi.org/10.1016/0550-3213(89)90130-2}{{\em Nucl. Phys. B}
  {\bfseries 323} (1989) 719--733}.

\bibitem{Fan:2021isc}
W.~Fan, A.~Fotopoulos, S.~Stieberger, T.~R. Taylor, and B.~Zhu, ``{Conformal
  Blocks from Celestial Gluon Amplitudes},''
  \href{http://arxiv.org/abs/2103.04420}{{\ttfamily arXiv:2103.04420
  [hep-th]}}.

\bibitem{Crawley:2021ivb}
E.~Crawley, N.~Miller, S.~A. Narayanan, and A.~Strominger, ``{State-Operator
  Correspondence in Celestial Conformal Field Theory},''
  \href{http://arxiv.org/abs/2105.00331}{{\ttfamily arXiv:2105.00331
  [hep-th]}}.

\bibitem{Banerjee:2020zlg}
S.~Banerjee, S.~Ghosh, and P.~Paul, ``{MHV Graviton Scattering Amplitudes and
  Current Algebra on the Celestial Sphere},''
  \href{http://arxiv.org/abs/2008.04330}{{\ttfamily arXiv:2008.04330
  [hep-th]}}.

\bibitem{Banerjee:2020vnt}
S.~Banerjee and S.~Ghosh, ``{MHV Gluon Scattering Amplitudes from Celestial
  Current Algebras},'' \href{http://arxiv.org/abs/2011.00017}{{\ttfamily
  arXiv:2011.00017 [hep-th]}}.

\bibitem{Pate:2019lpp}
M.~Pate, A.-M. Raclariu, A.~Strominger, and E.~Y. Yuan, ``{Celestial Operator
  Products of Gluons and Gravitons},''
  \href{http://arxiv.org/abs/1910.07424}{{\ttfamily arXiv:1910.07424
  [hep-th]}}.

\end{thebibliography}\endgroup

\end{document}